%% file: RadRecombPubl5.tex
\documentclass{article}
\usepackage{amsmath,amsbsy,amsfonts}
\usepackage[dvips]{graphics,color}
\usepackage{graphicx}

\begin{document}
\input{Commands}
\input{FigurecommandsV.tex}
\input{FigGreen}
\setlength{\unitlength}{1cm}
\newcommand{\red}{\textcolor{red}}

\newcommand{\calT}{{\cal T}}
\newcommand{\calTT}{{\widetilde{\cal T}}}
\newcommand{\calGT}{{\widetilde{\cal G}}}
\renewcommand{\calR}{{\cal R}}
\newcommand{\kp}{\kappa}
\newcommand{\GA}{\Gamma}
\newcommand{\GP}{\Gamma_P}
\newcommand{\PI}{\,P{\hspace{-3mm}-}}

\title{QED effects in scattering processes involving atomic bound states: Radiative recombination}
\author{Ingvar Lindgren and Sten Salomonson, Department of Physics,\\University of Gothenburg, 412 96 Gothenburg, Sweden, \\and Johan Holmberg, Physikalisches Institut, Universit\Št Heidelberg, \\69120 Heidelberg, Germany}
\maketitle

\abstract
The standard S-matrix formulation cannot generally be used in the treatment of atomic scattering processes, involving bound-state QED effects, due to the special type of singularity that can here appear. This type of singularity can be handled by means of  methods designed for structure calculations. It is essentially a consequence of the optical theorem that similar techniques can be applied also in scattering processes. The optical theorem for free particles gives a relation between the effective Hamiltonian and the cross section, a relation that is valid also when bound states are present. We  have found that the method with the Covariant-evolution-operator/Green's operator that we have developed primarily for structure problems can here be applied in a rather straightforward manner. The new procedure is demonstrated for the case of radiative recombination.

\section{Introduction}
There is presently an increasing interest in studying the effects of quantum electrodynamics (QED) on various dynamical processes. At GSI in Darmstadt high-energy collisions, involving highly charged ions, have been intensively studied, particularly the process known as \it{radiative electron capture} (REC), where a loosely bound electron is being captured by the projectile ion under the emission of a photon~\cite{REC07}. A closely related process is \it{radiative recombination}, where an electron in the continuum is being captured by the target ion also under the emission of a photon.

When bound states are involved, a certain type of singularity might appear, in structure calculations referred to as \it{model-space contributions}. These cannot be handled with the standard S-matrix formalism, but several methods have been developed for dealing with that in structure calculations. One procedure is the S-matrix in combination with the Sucher energy formula, which contains counterterms to eliminate the singularities~\cite{MPS98,Su57}. Another procedure is the Two-Times Green's function, developed by Shabaev and coworkers in St Petersburg~\cite{Shab02}. A third method is the Covariant Evolution Operator method, developed by the Gothenburg group~\cite{LSA04,ILBook11}. The evolution operator can be singular, and eliminating the singularities leads to what is referred to as the \it{Green's operator}, $\calG$, due to its analogy with the Green's function. In structure  calculations this procedure also yields information about the wave function or wave operator.

From the optical theorem for free particles~\cite{PS95} it follows that the imaginary part of the effective Hamiltonian is closely related to the scattering cross section - a relation that is valid also when bound states are present. This implies that methods originally designed for structure problems can be used also for scattering problems. Recently, Shabaev et al. have studied the QED corrections to the radiative recombination process, involving a bare nucleus, by applying the two-times Green's function~\cite{ShabRR00, ShabRR94}. An alternative procedure is the Green's operator, which when running over all times is also closely related to the effective Hamiltonian. We have found that this leads to a particularly simple and direct procedure. We have applied this technique to the radiative recombination process, studied by Shabaev et al., and found excellent agreement.

\section{Scattering processes}
Our treatment of the scattering process will be based upon the effective Hamiltonian and its relation to the scattering cross section. We shall first derive this relation from the optical theorem for free particles, a relation that holds also when bound states are present. This procedure will be applied to the case of radiative recombination.

\subsection{Scattering of free particles. Optical theorem}

The scattering process for free particles is normally described by means of the S-matrix. 
The scattering amplitude $\tau$ is related to the S-matrix by
\begin{equation}\label{ScattS}
  \bra{q}S\ket{p}=2\pi\im\delta(E_p-E_q)\,{\tau}(p\rarr q).
\end{equation}

An important tool in the study of scattering processes is the \ul{optical theorem} (see, for instance, Peskin and Schr\šder~\cite[p.230]{PS95}), which can be shown as follows.  We introduce 
\begin{equation}\label{T}
  S=1+\im T\,;\qquad S\dagg=1-\im T\dagg.
\end{equation}
Since $S$ is unitary, we have
\begin{equation}\label{SS}
  1=SS\dagg=1+\im(T-T\dagg)+T\dagg T,
\end{equation}
or 
\begin{equation}\label{SS1}
  -\im(T-T\dagg)=T\dagg T.
\end{equation}
We consider a diagonal element of this equation and insert a complete set of intermediate states on the rhs
\begin{equation}\label{SS2}
  -\im\bra{p}T-T\dagg\ket{p}=\sum_q \bra{p}T\dagg\ket{q}\bra{q}T\ket{p}
  =\sum_q \bra{q}T\ket{p}^*\bra{q}T\ket{p}.
\end{equation}
This gives
\begin{equation}\label{SS3}
  2Im{\bra{p}T\ket{p}}=\sum_q \bra{q}T\ket{p}^*\bra{q}T\ket{p}=\sum_q \big|\bra{q}T\ket{p}\big|^2,
\end{equation}
and using the relation \eqref{ScattS}
\begin{equation}\label{OT}
  \boxed{2Im{\bra{p}T\ket{p}}=\sum_q \big|\bra{q}T\ket{p}\big|^2=
  \sum_q \Big|2\pi\delta(E_p-E_q)\tau(p\rarr q)\Big|^2.}
\end{equation}
This implies that \bfit{the imaginary part of the forward scattering amplitude is proportional to the total cross section, which is the \ul{optical theorem}}. 

The forward scattering amplitude becomes imaginary when an intermediate state goes on-shell, and
this part can therefore be obtained by considering possible cases for this.
Cutkosky~\cite[p. 236]{PS95} has given the following rules for applying the optical theorem to a Feynman diagram:
\begin{itemize}
  \item Cut through all diagrams in all possible ways such that the cut propagators can simultaneously be put on shell; 
   \item For each cut, replace $1/(p^2-m^2+\ime)$ by $-2\pi \im\delta(p^2-m^2)$ in each propagator and then perform the loop integrals;
   \item Sum the imaginary contributions for all possible cuts.
\end{itemize}

\subsection{Scattering involving bound particles}

The standard S-matrix formalism cannot describe the scattering process, when there is a degeneracy involving a \ul{discrete} state. The reason is that this leads to a singularity, which requires a special treatment (model-space contribution, MSC). 

It is shown in the Appendix \eqpref{SHeff} that the diagonal element of the forward scattering amplitude is closely related to the \it{effective Hamiltonian}
\begin{equation}\label{THeff}
  PTP=-2\pi\delta(E_\rm{in}-E_\rm{out})W,
\end{equation}
where $W$ is the \it{effective iteration} $W=H\eff-PH_0P$ and $H\eff$ is the effective Hamiltonian. $P$ is the projection operator for the model space, degenerate with the initial state.
The optical theorem \eqref{OT} can then be expressed
\begin{equation}\label{OTB}
   \boxed{2Im{\bra{p}-H\eff\ket{p}}=  \sum_q2\pi\delta(E_p-E_q) \Big|\tau(p\rarr q)\Big|^2,}
\end{equation}
since the model Hamiltonian $H_0$ has no imaginary part. This is valid also when bound states are present, but the effective Hamiltonian then has to be evaluated in a different way.

From the expression for the differential cross section
\begin{equation}
  \dif\sigma=\frac{(2\pi)^4}{v_i}\,\delta(E_p-E_q)\,\big|\tau(p\rarr q)\big|^2\dif\bs{k},
\end{equation}
where $v_i$ is the velocity of the electron relative to the nucleus, we obtain the the important \bfit{relation between the effective Hamiltonian and the scattering cross section}
\begin{equation}
  \boxed{\dif\sigma=\frac{(2\pi)^3}{v_i}\,2Im(-H\eff)\,\dif\bs{k}.}
\end{equation}

In order to evaluate $2Im(-H\eff)$, we can use the same Cutkosky rules as before, slightly modified:
\begin{itemize}
  \item Make one cut in all diagrams of the effective Hamiltonian in all possible ways so that the cut state can be degenerate with the initial and final states;
   \item For each cut, replace the singularity $1/(A+\ime)$ by $-2\pi \im\delta(A)$ and for the remaining degeneracies evaluate the model-space contributions;   
   \item Sum all imaginary contributions.
\end{itemize}

\section{Radiative recombination}
\begin{figure}
\begin{center}
  \begin{picture}(2,3.5)(-1,0)\setlength{\unitlength}{0.8cm}
     \put(0,2){\photonENE{k}{}{}}
     \put(0,0.5){\lineD{1.5}}\put(0,2){\elline{2}{a}{}{}}
  \put(-0.5,1){\makebox(0,0){$p$}}
\end{picture}
\renewcommand{\normalsize}{\footnotesize}
    \caption{Lowest-order process in radiative recombination. The solid line represents an electron in  a bound state and a thin double-line a "quasi-free" electron in the continuum, moving in the nuclear potential.}
    \label{Fig:Exp}
\end{center}
\end{figure}
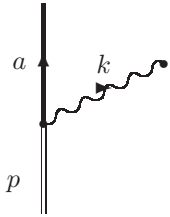

We shall now apply the Green's-operator procedure to the process of radiative recombination.\footnote{For details concerning the Covariant evolution operator and the Green's operator we refer to the Appendix and to ref.~\cite{ILBook11}.} In this process  an incoming electron is captured by an ion (in this particular case assumed to be a naked nucleus), as illustrated in Fig. \ref{Fig:Exp}.

\subsection{Lowest  order}
\begin{figure}
\begin{center}
 \begin{picture}(5,5)(-1,0)\setlength{\unitlength}{1cm}
   \put(0,0.5){\lineD{1}} \put(0,3.5){\lineD{1}}
  \put(0,1.5){\Elline{2}{0.7}{n=a\hsp}{\eps_n}{}}
   \put(0,2.5){\setlength{\unitlength}{1cm}\ElSE{}{}{}} 
  \put(-1,2.5){\line(1,0){2.8}}\put(2.7,2.5){\makebox(0,0){$\LineH{1}$}}
  \put(3.6,2.65){\setlength{\unitlength}{2cm}\makebox(0,0){$\VectorR$}}
  \put(1.2,2.2){\makebox(0,0){$k$}}
  \put(-0.5,0.75){\makebox(0,0){$p$}} \put(-0.5,4.255){\makebox(0,0){$p$}}
   \put(1.2,0.75){\makebox(0,0){$\eps_p=\eps_n+c\kp$}}
   \put(2.7,2.5){\makebox(0,0){$\LineH{1}$}}
\end{picture}
 \begin{picture}(2,4.5)(-1,0)\setlength{\unitlength}{0.6cm}
   \put(0,6.5){\lineD{1.5}}\put(0,4.5){\elline{2}{a}{}{}}
  \put(2,5.5){\photonWNW{k}{}{}}
    \put(0,2){\photonENE{k}{}{}}
     \put(0,0.5){\lineD{1.5}}\put(0,2){\elline{2}{a}{}{}}
  \put(-0.5,1){\makebox(0,0){$p$}}\put(-0.5,7.3){\makebox(0,0){$p$}}
   \put(0.7,1){\makebox(0,0){$\eps_p$}}
\end{picture}
\renewcommand{\normalsize}{\footnotesize}
    \caption{Applying the optical theorem in lowest order, left: forward scattering amplitude, right: after the cut.}
    \label{Fig:Lowest}
\end{center}
\end{figure}
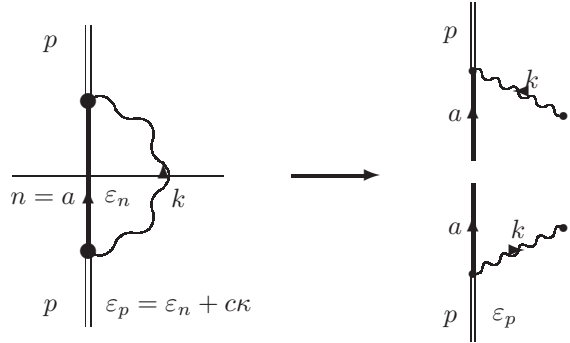

In lowest order the forward scattering amplitude is represented by the Feynman diagram to the left in Fig. \ref{Fig:Lowest}. The Green's operator is in this case identical to the S-matrix and given by means of the Feynman rules for the S-matrix~\cite[Ch. 7]{MS84}, \cite[App. H]{ILBook11}
\begin{equation}
   \bra{p}S\ket{p}=\bra{p}\calG(\infty,-\infty)\ket{p}=2\pi\delta(E_\rm{in}-E_\rm{out})\bra{p}\im A \im\GA \im A\ket{p},
\end{equation}
where  $A$ stands for the photon interaction $A=ec\alpha^\mu A_\mu$ and 
\begin{equation}
  \GA=\GA(\eps_p)=\frac{\ket{n+k}\bra{n+k}}{\eps_p-\eps_n-c\kp+\ime}
 \end{equation}
is the resolvent.  (Normally we employ the sum convention.) The $k$ vector is $k=(c\kp,-\textbf{k})$. The photon-field operators are assumed to be contracted.

Using the relation in the Appendix \eqref{SHeff}
\begin{equation}  \label{Heff2}
  P\im\calG(\infty,-\infty)P=2\pi\delta(E_\rm{in}-E_\rm{out})W,
\end{equation}
we then get
\begin{equation}  
  \bigbra{p}-W\bigket{p}=-\bigbra{p}A\GA A\bigket{p}=-\bigbra{p}A \frac{\ket{n+k}\bra{n+k}}{\eps_p-\eps_n-c\kp+\ime}A\bigket{p}.
\end{equation}
This has a singularity, when $\eps_p=\eps_n+c\kp$
\begin{equation}\label{S1sing}
  W\rarr\bra{p}A \GP A\ket{p}
\end{equation}
with $\GP=P\GA$, which lies in a \bfit{continuum}, since the photon energy is not fixed. Therefore, the integral leads to a principal integral and half a pole contribution $\PI-\pi\im\,\delta(\eps_p-\eps_n-c\kp)$. Then
\begin{equation}
  2Im \bra{p}-H\eff\ket{p}=2\pi\delta(\eps_p-\eps_a-c\kp)\bra{p}A\ket{q}\bra{q}A\ket{p},
  \end{equation}
since $H_0$ does not have any imaginary part. $\ket{q}$ stands here for the intermediate state $\ket{n,k}$, where $n$ is a bound state and $k$ represents the photon.
The renormalization of the self-energy leads to a counter mass term, which, however, does not contribute to the imaginary part.  

We have now found an expression that is proportional to the lowest-order the scattering cross section. Next we shall consider higher-order contributions.

\subsection{Self-energy insertion on the bound state}
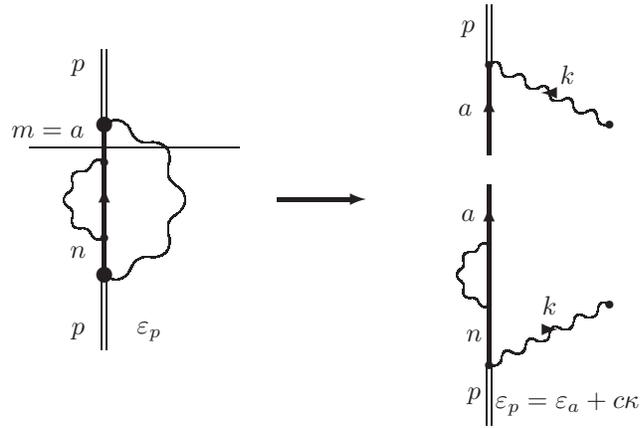
\begin{figure}
\begin{center}
 \begin{picture}(5,4.5)(0,-0.5)\setlength{\unitlength}{1cm}
  \put(0,0.5){\lineD{1}} \put(0,3.5){\lineD{1}}
  \put(0,1.5){\elline{2}{}{}{}}
    \put(-1,3.2){\line(1,0){2.8}}\put(2.3,2.5){\makebox(0,0){$\LineH{1}$}}
    \put(3.2,2.65){\setlength{\unitlength}{2cm}\makebox(0,0){$\VectorR$}}
    \put(0,2.5){\setlength{\unitlength}{1cm}\ElSEG{}{}{}} 
     \put(0,2.5){\setlength{\unitlength}{0.5cm}\ElSELG{}{}{}} 
  \put(-0.35,0.75){\makebox(0,0){$p$}} \put(-0.35,4.255){\makebox(0,0){$p$}}
   \put(0.6,0.75){\makebox(0,0){$\eps_p$}}
    \put(-0.75,3.4){\makebox(0,0){$m=a$}} \put(-0.35,1.8){\makebox(0,0){$n$}}
\end{picture}
 \begin{picture}(3,6)(0,0)\setlength{\unitlength}{0.8cm}
    \put(0,6){\lineD{1}}\put(0,4.5){\elline{1.5}{a}{}{}}
 \put(2,5){\photonWNW{k}{}{}}
 \put(-0.35,6.6){\makebox(0,0){$p$}}
  \put(0,0){\lineD{1}}\put(0,1){\Elline{3}{2.5}{a}{}}
    \put(0,1){\photonENE{k}{}{}}
 \put(0,2.5){\setlength{\unitlength}{0.4cm}\ElSELG{}{}{}} 
  \put(-0.25,0.5){\makebox(0,0){$p$}}\put(-0.25,1.5){\makebox(0,0){$n$}}
 \put(1.3,0.35){\makebox(0,0){$\eps_p=\eps_a+c\kp$}}
\end{picture}
\renewcommand{\normalsize}{\footnotesize}
    \caption{Self-energy on the bound state - cut at upper state.}
    \label{Fig:SEBound}
\end{center}
\end{figure}
\begin{figure}
\begin{center}
 \begin{picture}(5,5)(0,-0.5)\setlength{\unitlength}{1cm}
  \put(0,0.5){\lineD{1}} \put(0,3.5){\lineD{1}}
  \put(0,1.5){\elline{2}{}{}{}}
    \put(-1,1.8){\line(1,0){2.8}}\put(2.3,2.5){\makebox(0,0){$\LineH{1}$}}
    \put(3.2,2.65){\setlength{\unitlength}{2cm}\makebox(0,0){$\VectorR$}}
    \put(0,2.5){\setlength{\unitlength}{1cm}\ElSEG{}{}{}} 
     \put(0,2.5){\setlength{\unitlength}{0.5cm}\ElSELG{}{}{}} 
  \put(-0.35,0.75){\makebox(0,0){$p$}} \put(-0.35,4.255){\makebox(0,0){$p$}}
   \put(0.3,0.75){\makebox(0,0){$\eps_p$}}
    \put(-0.4,3.4){\makebox(0,0){$m$}} \put(-0.75,1.65){\makebox(0,0){$n=a$}}
\end{picture}
 \begin{picture}(2,4.)(0,0)\setlength{\unitlength}{0.8cm}
  \put(0,6){\lineD{1}}\put(0,3){\Elline{3}{0.5}{a}{}}
    \put(2,5){\photonWNW{k}{}{}}
 \put(0,4.5){\setlength{\unitlength}{0.4cm}\ElSELG{}{}{}} 
  \put(-0.25,6.5){\makebox(0,0){$p$}}
   \put(-0.3,5.5){\makebox(0,0){$n$}}
       \put(0,0.5){\lineD{1}}\put(0,1.5){\elline{1.2}{a}{}{}}
        \put(0,1.5){\photonENE{k}{}{}}
  \put(-0.4,1){\makebox(0,0){$p$}}
   \put(0.4,1){\makebox(0,0){$\eps_p$}}
\end{picture}
\renewcommand{\normalsize}{\footnotesize}
    \caption{Self-energy on the bound state - cut at lower state.}
    \label{Fig:SEBound2}
\end{center}
\end{figure}
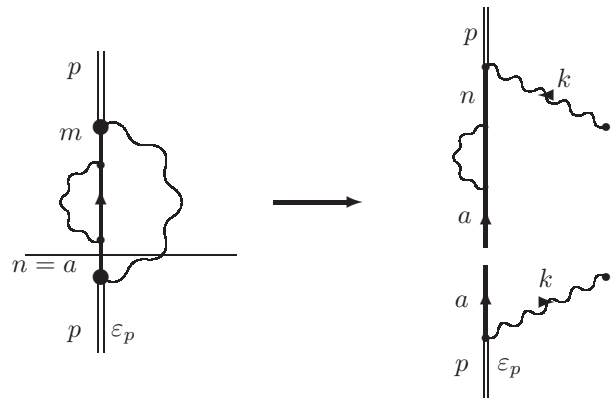

We begin the study of higher-order effects by considering the case when there is a self-energy insertion in the bound state. The  forward scattering amplitude is represented by the Feynman diagram in Figs \ref{Fig:SEBound} and \ref{Fig:SEBound2}, left, and we shall evaluate the effective Hamiltonian \eqref{Heff2}.
The evolution operator, which is singular, is given by 
  \begin{equation}
   \bra{p}U\ket{p}=2\pi\delta(E_\rm{in}-E_\rm{out})\,\bra{p}\im A \im\GA(-\im)\Sigma\im\GA\im A\ket{p},
  \end{equation}
where  $\Sigma$ stands for self-energy insertion
  \begin{equation}
   \Sigma=\Sigma(\eps_a)=\Sigma(\eps_p-c\kp).
  \end{equation}
The corresponding part of $ \bra{p} -H\eff\ket{p}$ is
 \begin{equation}
   \bra{p} -H\eff\ket{p}: \;-\bra{p} A\GA\,\Sigma\GA A\ket{p}.
  \end{equation}
 This has a regular part
   \begin{equation}\label{UReg}
    - \bra{p} A\GQ\,\Sigma\GQ A\ket{p},
  \end{equation}
 where the model-space states are eliminated by the reduced resolvent \eqref{ResolQ}. This does not contribute to the scattering cross section, since only the imaginary part of the forward scattering amplitude does.
 
 The imaginary part of the forward scattering amplitude is given by the singularities.
 One singularity can be treated as in first order, leading to
   \begin{equation}
 -\bra{p}A \GP\Sigma\GA A\ket{p}= -\,\PI+\im\pi\delta(\eps_p-\eps_a-c\kp)\bra{p}A\ket{q}\bra{q}\Sigma\GA A\ket{p},
  \end{equation}
   \begin{equation}
  -\bra{p}A \GA\Sigma\GP A\ket{p}= -\,\PI+\im\pi\delta(\eps_p-\eps_a-c\kp)\bra{p}A\GA\Sigma\ket{q}\bra{q} A\ket{p},
  \end{equation}
  where $\GP=P\GA$.
The photon energy is now \bfit{fixed}, and the other singularity lies in a \bfit{discrete} environment, and therefore leads to a model-space contribution, MSC (see Appendix)
\begin{equation}
  -\bra{p}A \GP\Sigma\GA A\ket{p} \rarr \im\pi\delta(\eps_p-\eps_a-c\kp)\Biggbra{p}\Pd{\calE}\Big(A\bigket{q}\bigbra{q}\Sigma(\calE-c\kp)\Big)_{\calE=\eps_p}\bigket{q}\bigbra{q} A\Biggket{p},
 \end{equation}
\begin{equation}
   -\bra{p}A \GA\Sigma\GP A\ket{p}\rarr \im\pi\delta(\eps_p-\eps_a-c\kp)\Biggbra{p}\Big(\pd{A}\Big)_{\calE=\eps_p}\bigket{q}\bigbra{q}\Sigma\bigket{q}\bigbra{q} A\Biggket{p}.
 \end{equation}
 These two contributions yield the imaginary part of the effective Hamiltonian \eqref{Heff2} and hence
\begin{eqnarray}\label{CrossSelf}
  &&2Im \bra{p}-H\eff\ket{p}=2\pi\delta(\eps_p-\eps_a-c\kp)\nn
 &\times& \Biggbra{p}A\ket{q}\bigbra{q}\Sigma\GQ A+A\GQ\Sigma\bigket{q}\bigbra{q}A+ 
  \Pd{\calE}\Big(A\bigket{q}\bigbra{q}\Sigma\Big)_{\calE=\eps_p}\bigket{q}\bigbra{q}A\nn
  &+&\Big(\pd{A}\Big)_{\calE=\eps_p}\bigket{q}\bigbra{q}\Sigma\bigket{q}\bigbra{q} A\Biggket{p}.
\end{eqnarray}
This is the corresponding contribution to the cross section in the case with a self-energy insertion on the bound state line.

The terms with the derivative of the photon interaction $ec\alpha^\nu A_\nu$ correspond to the correction of the photon energy due to the modification of the bound-state electron energy, caused by first-order QED corrections, given by Shabaev et al.~\cite[Eq. (34)]{ShabRR00}.

\subsection{Vertex correction}

\begin{figure}
\begin{center}\setlength{\unitlength}{0.9cm}
 \begin{picture}(6,5)(0,0)
  \put(0,0){\lineD{1}} \put(0,4){\lineD{1}}
      \put(-1,3.3){\line(1,0){2.8}}\put(2.3,2.5){\makebox(0,0){$\LineH{1}$}}
    \put(3.2,2.65){\setlength{\unitlength}{2cm}\makebox(0,0){$\VectorR$}}
  \put(0,1){\elline{3}{}{}{}}
    \put(0,3){\setlength{\unitlength}{0.8cm}\ElSEG{}{}{}} 
   \put(0,2){\setlength{\unitlength}{0.8cm}\ElSELG{}{}{}} 
  \put(-0.35,0.5){\makebox(0,0){$p$}} \put(-0.35,4.5){\makebox(0,0){$p$}}
   \put(0.4,0.5){\makebox(0,0){$\eps_p$}}
    \put(-0.75,3.5){\makebox(0,0){$m=a$}} \put(-0.3,1.5){\makebox(0,0){$n$}}
  \put(-0.35,2.1){\makebox(0,0){$\nu$}}
\end{picture}
 \begin{picture}(3,5)(0,0)\setlength{\unitlength}{0.7cm}
     \put(0,6){\lineD{1}}\put(0,4.5){\elline{1.5}{a}{}{}}
 \put(2,5){\photonWNW{k}{}{}}
 \put(-0.35,6.6){\makebox(0,0){$p$}}
  \put(0,0){\lineD{1}} 
  \put(0,1){\elline{3.25}{}{}{}}
    \put(0,1.75){\photonENE{k}{}{}} 
   \put(0,2){\setlength{\unitlength}{0.8cm}\ElSELG{}{}{}} 
  \put(-0.35,0.25){\makebox(0,0){$p$}} 
   \put(0.4,0.2){\makebox(0,0){$\eps_p$}}
    \put(-0.35,3.7){\makebox(0,0){$a$}} \put(-0.3,1.35){\makebox(0,0){$n$}}
   \put(-0.3,1.75){\makebox(0,0){$\nu$}}
\end{picture}
\renewcommand{\normalsize}{\footnotesize}
    \caption{Scattering with a vertex correction - cut at the upper state.}
    \label{Fig:Vertex}
\end{center}
\end{figure}
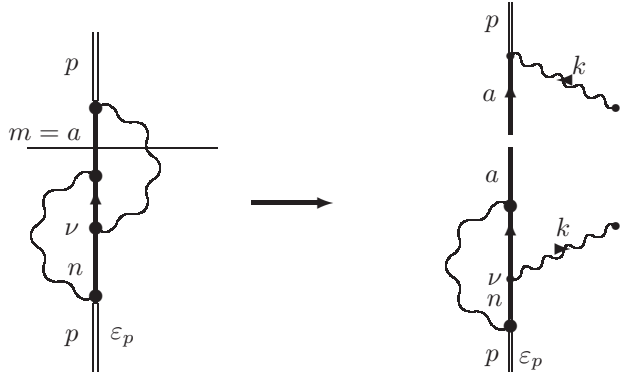

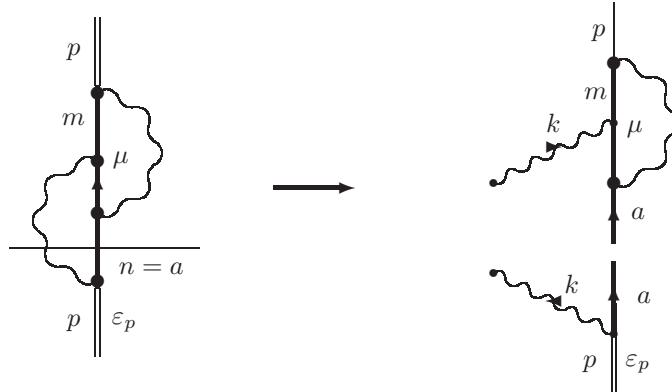
\begin{figure}
\begin{center}\setlength{\unitlength}{0.9cm}
 \begin{picture}(5.5,5)(0,0)
  \put(0,0){\lineD{1}} \put(0,4){\lineD{1}}
      \put(-1.3,1.6){\line(1,0){2.8}}\put(2.6,2.5){\makebox(0,0){$\LineH{1}$}}
    \put(3.5,2.65){\setlength{\unitlength}{2cm}\makebox(0,0){$\VectorR$}}
  \put(0,1){\elline{3}{}{}{}}
    \put(0,3){\setlength{\unitlength}{0.8cm}\ElSEG{}{}{}} 
   \put(0,2){\setlength{\unitlength}{0.8cm}\ElSELG{}{}{}} 
  \put(-0.35,0.5){\makebox(0,0){$p$}} \put(-0.35,4.5){\makebox(0,0){$p$}}
   \put(0.4,0.5){\makebox(0,0){$\eps_p$}}
    \put(-0.35,3.5){\makebox(0,0){$m$}} \put(0.8,1.3){\makebox(0,0){$n=a$}}
  \put(0.35,2.9){\makebox(0,0){$\mu$}}
\end{picture}
\begin{picture}(2,4.)(-2,1)\setlength{\unitlength}{0.8cm}
  \put(0,6){\line(0,4){1}}\put(0,3){\Elline{3}{0.5}{}{a}}
    \put(-2,4){\photonENE{k}{}{}}
 \put(0,5){\setlength{\unitlength}{0.8cm}\ElSEG{}{}{}} 
  \put(-0.25,6.5){\makebox(0,0){$p$}}
   \put(-0.3,5.5){\makebox(0,0){$m$}}
       \put(0,0.5){\lineD{1.5}}\put(0,1.5){\elline{1.2}{}{a}{}}
      \put(0,1.5){\photonWNW{k}{}{}}
  \put(-0.4,1){\makebox(0,0){$p$}}
   \put(0.4,1){\makebox(0,0){$\eps_p$}}
    \put(0.35,4.9){\makebox(0,0){$\mu$}}
\end{picture}
\renewcommand{\normalsize}{\footnotesize}
    \caption{Scattering with a vertex correction - cut at the lower state.}
    \label{Fig:Vertex2}
\end{center}
\end{figure}

Next we consider the contribution to the scattering amplitude, when  there is
 a vertex correction as illustrated in Figs \ref{Fig:Vertex} and \ref{Fig:Vertex2} (left). The contribution to the forward scattering amplitude is\footnote{Sign conventions (see, for instance ref.~\cite{ILBook11}): $\Sigma(\eps_a)=\im\intd{z}\,\SF(\eps_a-z;\bx_2,\bx_1)\,I(z;\bx_2,\bx_1)$. $\Lambda^\mu(\eps_a,\eps_a)
   =-\im\alpha^\mu\intd{z}\SF(\eps_a-z;\bx_2,\bx_3)  \,\SF(\eps_a-z;\bx_3,\bx_1)   \,I(z;\bx_2,\bx_1)$. 
  $ I(z;\bx_1,\bx_2)=e^2c^2\alpha_1^\mu\alpha_2^\nu\DFmn(z;\bx_1,\bx_2)$.}
\begin{equation}\label{Vx1}
  2\pi\delta(E_\rm{in}-E_\rm{out})\,\bigbra{p}-\im\Lambda\,\im\GA\,\im A\bigket{p} \quad \rm{or}\quad  2\pi\delta(E_\rm{in}-E_\rm{out})\,\bigbra{p}-\im A\,\im\GA\,\im \Lambda\bigket{p}
   \end{equation}
with contractions between the field operators.  $\Lambda$ stands for the vertex-correction interaction
$\Lambda=ec\Lambda^\mu A_\mu$. The corresponding part of the effective interaction becomes
\begin{eqnarray}
  \bigbra{p}-W\bigket{p}&=&\bigbra{p}\Lambda\,\GA\, A\bigket{p} \quad \rm{or}\quad
  \bigbra{p}A\,\GA\, \Lambda\bigket{p}
   \end{eqnarray}

One singularity can be eliminated as before, leading to
   \begin{equation}
 \bra{p}\Lambda\GP A\ket{p}=\PI-\im\pi\delta(\eps_p-\eps_a-c\kp)\bra{p}\Lambda\ket{q}\bra{q}A\ket{p},
  \end{equation}
   \begin{equation}
 \bra{p}A \GP \Lambda\ket{p}=\PI-\im\pi\delta(\eps_p-\eps_a-c\kp)\bra{p}A\ket{q}\bra{q} \Lambda\ket{p}.
  \end{equation}
There are no MSC in this case. This leads to
\begin{equation}\label{CrossVert}
   2Im \bra{p}-H\eff\ket{p}=-2\im\pi\delta(\eps_p-\eps_a-c\kp)\Big[\bigbra{p}\Lambda\ket{q}\bra{q}A\ket{p}+\bra{p}A\ket{q}\bra{q} \Lambda\bigket{p}\Big].
  \end{equation}

Our result differs in the relative sign from that of Shabaev et al., which might be due to different sign conventions (see footnote).

\subsection{Self-energy insertion on the free-electron state}

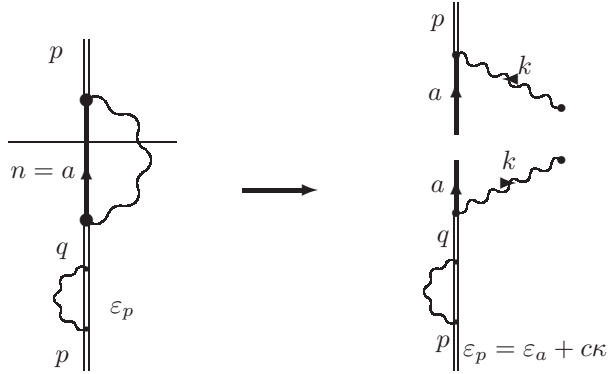
\begin{figure}
\begin{center}\setlength{\unitlength}{0.8cm}
 \begin{picture}(6,6)(0,-1)
     \put(-1.3,2.8){\line(1,0){2.8}}\put(2.6,2){\makebox(0,0){$\LineH{1}$}}
    \put(3.5,2.18){\setlength{\unitlength}{2cm}\makebox(0,0){$\VectorR$}}
  \put(0,-1){\lineD{2.5}} \put(0,3.5){\lineD{1}}
  \put(0,1.5){\Elline{2}{0.8}{n=a\hsp\;}{}{}}
    \put(0,2.5){\setlength{\unitlength}{0.8cm}\ElSEG{}{}{}} 
     \put(0,0.2){\setlength{\unitlength}{0.4cm}\ElSELG{}{}{}} 
  \put(-0.4,-0.8){\makebox(0,0){$p$}} \put(-0.4,1){\makebox(0,0){$q$}}
   \put(-0.5,4.255){\makebox(0,0){$p$}}
   \put(0.6,0){\makebox(0,0){$\eps_p$}}
\end{picture}
 \begin{picture}(3.5,4.5)(0,0)\setlength{\unitlength}{0.7cm}
    \put(0,6){\lineD{1}}\put(0,4.5){\elline{1.5}{a}{}{}}
 \put(2,5){\photonWNW{k}{}{}}
 \put(-0.35,6.6){\makebox(0,0){$p$}}
 \put(0,0){\lineD{3}}\ \put(0,3){\Elline{1}{0.5}{a}{}}
    \put(0,3){\photonENE{k}{}{}}
 \put(0,1.5){\setlength{\unitlength}{0.4cm}\ElSELG{}{}{}} 
  \put(-0.25,0.5){\makebox(0,0){$p$}}\put(-0.25,2.5){\makebox(0,0){$q$}}
    \put(1.5,0.35){\makebox(0,0){$\eps_p=\eps_a+c\kp$}}
\end{picture}
\renewcommand{\normalsize}{\footnotesize}
    \caption{Scattering with a self-energy correction on the free electron.}
    \label{Fig:FreeEl}
\end{center}
\end{figure}

When there is a self-energy insertion in the (quasi)free electron state, the forward scattering amplitude is represented by the Feynman diagram in Fig. \ref{Fig:FreeEl} (left), corresponding to the evolution operator
\begin{equation}\label{Fr1a}
  \bra{p}U\ket{p}= 2\pi\delta(E_\rm{in}-E_\rm{out})\,\Bigbra{p}\im A\im\GA\,\im A\im\GA(-\im)\Sigma\Bigket{p},\end{equation}
  which leads to
  \begin{equation}
  \bra{p}-W\ket{p}= -\Bigbra{p}A\GA\,A\GA\Sigma\Bigket{p}.
\end{equation}

The leftmost singularity corresponds to
\begin{equation}
  -\Bigbra{p}A\GP\,A\GA\Sigma\Bigket{p}= -\PI+\im\pi\delta(\eps_p-\eps_a-c\kp)\bra{p}A\ket{q}\bra{q}A\GA\Sigma\ket{p} 
\end{equation}
and
\begin{equation}
  \Bigbra{p}2Im(-H\eff)\Bigket{p}=2\im\pi\delta(\eps_p-\eps_a-c\kp)\bra{p}A\ket{q}\bra{q}A\GA\Sigma\ket{p} 
\end{equation}

  The other singularity does not correspond to any process of physical interest.
There is no MSC here.

There is also an inverted diagram with a self-energy on the outgoing line. This lead to
\begin{eqnarray}\label{S5}
 &&\mHsp\bra{p}2Im (-H\eff)\ket{p}=2\pi\delta(\eps_p-ck-\eps_a)
 \Biggbra{p}A\bigket{q}\bigbra{q}A\GA\,\Sigma+\Sigma\GA A\ket{q}\bra{q}A\Biggket{p}\nn
 &&\mHsp=2\pi\delta(\eps_p-ck-\eps_a)\Biggbra{p}A\ket{q}\bra{q}A\frac{\ket{n}\bra{n}}{\eps_p-\eps_n+\ime}\,\Sigma+\Sigma\frac{\ket{n}\bra{n}}{\eps_p-\eps_n+\ime}A\ket{q}\bra{q}A\Biggket{p}.\nn
\end{eqnarray}
Since the states $n$ are here continuous, the singularity leads to a principal integral and half a pole, of which only the latter contributes to the imaginary part.

\pagebreak
\section{Scattering amplitude}
By summing all contributions of $2Im \bra{p}\calT\ket{p}$ for this particular process, we have according to \eqref{OTB}
\begin{equation}
   2Im \bra{p}-H\eff\ket{p}=2\pi\delta(E_p-E_q)\big|\tau(p\rarr q)\big|^2.
\end{equation}
All contributions above are of the form
\begin{equation}
  2Im \bra{p}-H\eff\ket{p}=2\pi\delta(E_p-E_q)\bigbra{p}....\bigket{p},
\end{equation}
which leads to the relation
\begin{equation}\label{AmpSq}
   \bigbra{p}....\bigket{p}=\big|\tau(p\rarr q)\big|^2.
\end{equation}

In the present case we do not have an explicit expression for the scattering amplitude as in the free case \eqref{ScattS}. Instead, if we want to have an expression for the amplitude, this has to be extracted from the cross section \eqref{AmpSq}. 

From the expressions derived above we have
\begin{eqnarray}\label{ScAmpl}
  &&\big|\tau(p\rarr q)\big|^2=\Bigbra{p}A\Bigket{q}\nn
  &&\mhsp\times\Biggbra{q}A+\Sigma\GQ A+\Big(\pd{\Sigma}\Big)_{\calE=\eps_p}\bigket{q}\bigbra{q}A-\Lambda+
  A\frac{\ket{n}\bra{n}}{\eps_p-\eps_n+\ime}\,\Sigma\Biggket{p}\nn
  &&\mHsp+\Biggbra{p}A\GQ\Sigma+2\Big(\pd{A}\Big)_{\calE=\eps_p}\bigket{q}\bigbra{q}\Sigma-\Lambda +\Sigma\frac{\ket{n}\bra{n}}{\eps_p-\eps_n+\ime}A\Biggket{q}\Bigbra{q}A\Bigket{p},\Hsp
 \end{eqnarray}
 which is of the form
 \begin{eqnarray}
   \big|\tau(p\rarr q)\big|^2=\bigbra{p}A\bigket{q}\bigbra{q}A+X\bigket{p}+
   \bigbra{p}Y\bigket{q}\bigbra{q}A\bigket{p}.
\end{eqnarray}
This leads to the approximate amplitude
\begin{equation}
   \tau(p\rarr q)\approx\bigbra{q}A+\halfS(X+Y\dagg)\Bigket{p},
\end{equation}
and
\begin{eqnarray}
 && \tau(p\rarr q)\approx\Biggbra{q}A+\Sigma\GQ A+\Sigma\bigket{q}\bigbra{q}\Big(\pd{A}\Big)_{\calE=\eps_p}\nn
&+&\half\Big(\pd{\Sigma}\Big)_{\calE=\eps_p}\bigket{q}\bigbra{q}A-\Lambda+A\frac{\ket{n}\bra{n}}{\eps_p-\eps_n+\ime}\,\Sigma\Biggket{p}.
 \end{eqnarray}
This agrees with the result of Shabaev et al.~\cite{ShabRR00} (apart from sign difference for the vertex part, which might be due to different sign convention).

The self-energies and vertex corrections that are not affected by the cuts have to be properly renormalized. For the effects affected by the cut the renormalization does not contribute to the imaginary part. 

It can be noted that the derivative of the self-energy appears only once in the cross-section, which explains the factor of one half in the amplitude. The derivative of the photon energy, on the other hand, appears twice and hence appears with the factor of unity in the amplitude.
  
The effect of the vacuum polarization is not included here but can be evaluated in very much the same way, and in doing so, we find also here agreement with the result of Shabaev et al.
    
\section{Summary and conclusions}
We have here demonstrated that the procedure of the covariant-evolution operator/Green's operator, originally developed for structure calculations, can also be readily applied in dynamical processes involving bound particles. This is here demonstrated particularly for the process of radiative recombination, but the procedure should be applicable also to other dynamical processes. Due to the presence of bound states in the process, the standard S-matrix procedure is not applicable. The results we obtain are in excellent agreement with those obtained by Shabaev et al., using the two-times Green's function, also originally developed for structure calculations. Our procedure is based upon the effective Hamiltonian, which according to the optical theorem is closely related to the scattering cross section.

\section*{Acknowledgements}
The authors wish to acknowledge Andrey Surzhykov and Anton Artemyev, as well as Eva Lindroth and Vladimir Shabaev, for stimulating discussions. J.H. acknowledges support from the Helmholtz Association and GSI under the project VH-NG-421.

\section*{Appendix. The covariant evolution operator and the Green's operator}

\subsection*{S-matrix}
We shall first consider the S-matrix, which relates the final and initial states in a dynamical process
\begin{equation}\label{Sdef}
    \Phi(t=+\infty)=S\,\Phi(t=-\infty)
\end{equation}
and is related to the standard evolution operator by
\begin{equation}\label{S}
    S=\U(\infty,-\infty).
\end{equation}
The S-matrix can be expanded as~\cite[Eq. 6.23]{MS84}, \cite[Eq. (4.3)]{ILBook11}
 \begin{equation}   \label{Sexp}
  S =\sum_{n=0}^{\infty} \frac{1}{n!}\Big(\frac{-\im}{c}\Big)^n
\int \dif x^4_{1} \ldots \int \dif x^4_{n}\; T\big[\calH(x_{1})
\ldots \calH(x_{n}) \big]\, \me^{-\gamma(|t_{1}|+|t_{2}| \ldots
|t_{n}|)},\Hsp
\end{equation}
integrated over all space and time. Here, $\ga$ is an adiabatic-damping parameter and
\begin{equation}\label{HI}
    \calH(x)=-\hpsi\dagg(x) ec\alpha^\mu A_{\mu}(x)\hpsi(x)
\end{equation}
represents the interaction between the electron (charge $-e$) and the electromagnetic field
  \begin{equation}
  A_\mu(x)=\frac{1}{\sqrt{2\kp c(2\pi)^3}}\,\sum_r\epsi_{\mu r}\Big[a_{k r}\me^{-\im k x}+a\dagg_{k r}\me^{\im k x}\Big],
  \end{equation}
 $k=(c\kappa,-\bs{k})$ being the covariant $k$ vector, $a\dagg_{k r},\;a_{kr}$ the creation and annihilation operators, and $\epsi_{\mu r}$ the components of the polarization vector, $r$ representing the polarization direction. $\hpsi(x),\;\hpsi\dagg(x)$ represent the electron-field operators.
  
The energy can be calculated by means of the Sucher formula~\cite{Su57}
\begin{equation}  \label{SPert}
  S=\sum_n S^{(n)}
\end{equation}
\begin{equation}  \label{ESucher}
  \Delta E=\lim_{\gamma\rarr0}\frac{\im\gamma}{2} \frac{\sum
  n\bra{\Phi}S^{(n)}\ket{\Phi}}{\bra{\Phi}S\ket{\Phi}}.
\end{equation}
This formula eliminates the singularity of the S-matrix that appears when there is an intermediate model-space state.

\subsection*{Covariant evolution operator and the Green's operator\,\footnote{See further refs~\cite{LSA04} and \cite[Chap. 6]{ILBook11}. }}

The S-matrix is relativistically covariant, which is not the case for the standard evolution operator for finite times. It can be modified, though, in such way that it becomes covariant for all times, leading to what we  refer to as the \it{covariant evolution operator} (CEO). In the single-particle case it can  be defined~
\begin{equation}\label{UCovDef}
   \U(t,t_0)=\dint\dif^3\bx\,\dif^3\bx_0\,\hpsi\dagg(x)
   \bigbra{0_\mH}T[\hpsi_\mH(x)\hpsi_\mH\dagg(x_0)]\bigket{0_\mH}\,\hpsi(x_0),
\end{equation}
using the \it{Heisenberg representation} and $T$ being the \it{Wick time-ordering} operator. In the \it{interaction representation} this leads to the expansion
\begin{eqnarray}\label{UCovExp}
    U(t,t_0)&=&\sum_{n=0}^{\infty}\frac{1}{n!}
     \dint\dif^3\bx\,\dif^3\bx_0\;\Big(\frac{-\im}{c}\Big)^n
    \int\dif^4x_1\cdots \int \dif^4x_n\nn
   &\times&\hpsi\dagg(x)\, T\Big[ \hpsi(x)\,{\calH}(x_1)\cdots{\calH}(x_n)
   \, \hpsi\dagg(x_0)\Big]\hpsi(x_0)\;\me^{-\gamma(|t_1|+|t_2|\cdots)},\Hsp
\end{eqnarray}
where $\ga$ is the adiabatic damping parameter. The operators are connected to form a one-body operator. In second order this becomes (leaving out the damping factor)
\begin{eqnarray}
   \U\tva(t,t_0)&=&\dint\dif^3\bx\,\dif^3\bx_0\,\hpsi\dagg(x)
  \im\SF(x,x_4)(-\im)V(x_4,x_3) \im\SF(x_3,x_2)\nn
  &\times&(-\im)V(x_2,x_1)\im\SF(x_1,x_0)\hpsi(x_0),
\end{eqnarray}
where $V$ is the perturbation. The factor of 1/2 is eliminated, if we consider only one of the two permutations of the perturbations.

If we let $t_0\rarr-\infty$, then
\begin{equation}\label{Ginf}
   \int\dif^3\bx_0\im\SF(x_1,x_0)\hpsi(x)\rarr\hpsi(x_1)
\end{equation}
this reduces to
\begin{equation}\label{U2R}
   \U\tva(t,-\infty)=\int\dif^3\bx\,\,\hpsi\dagg(x)
  \SF(x,x_4)\,V(x_4,x_3) \SF(x_3,x_2) \,V(x_2,x_1)\hpsi(x_1).
\end{equation}
After Fourier transforming the operators, this can be expressed
\begin{equation}\label{Ladd2}
   U\tva(t,-\infty)\PE=\me^{-\im t(\calE-H_0)}\GE V(\calE)\GE V(\calE)\PE,
\end{equation}
where $\PE$ is the projection operator for a part of the model space of energy $\calE$ and
\begin{subequations}
\begin{equation}\label{Resol}
   \SF(\calE)=\GE=\frac{1}{\calE-H_0}
\end{equation}
is the \it{resolvent operator}. Later we shall also need the \it{reduced resolvent operator}
\begin{equation}\label{ResolQ}
   \GQE=\frac{Q}{\calE-H_0},
\end{equation}
\end{subequations}
where $Q=1-P$ is the projection operator for the space outside the model space.
$H_0$ is the zeroth-order or model Hamiltonian that generates the zeroth-order or model functions.

A general single-particle ladder can be expressed
\begin{equation}\label{Un}
   U\n(t,-\infty)\PE=\me^{-\im t(\calE-H_0)}\GE V(\calE)\GE V(\calE)\GE V(\calE)\cdots\PE.
\end{equation}
This becomes (quasi)singular, when an intermediate or final state is 
(quasi) degenerate with the initial state. In order to remedy that we introduce the \bfit{Green's operator}, defined by the relation
\begin{equation}  \label{GreenMR}
  U(t-\infty)P=\calG(t,-\infty)\bdot P\U(0,-\infty)P,
\end{equation}
where the dot indicates that the Green's operator acts on the \ul{intermediate} model-space state.
The Green's operator can be shown to be \bfit{regular}.

We apply the definition \eqref{GreenMR} to the operator \eqref{Un} (with simplified notations)
\begin{equation}\label{GnDef}
   U\n(\calE)\PE=\calG\n(\calE) \PE+\calG\nett(\calE')\PEP U\ett(\calE)\PE+\calG\ntva(\calE')\PEP U\tva(\calE)\PE     +\cdots,
\end{equation}
noting that $U\nol(0)=1$. Solving for $\calG\n$, we have
\begin{equation}\label{Gn}
   \calG\n(\calE)\PE=U\n(\calE) \PE-\calG\nett(\calE')\PEP U\ett(\calE)\PE-\calG\ntva(\calE')\PEP U\ntva(\calE)\PE-\cdots,
\end{equation}
The negative terms are known as \bfit{counterterms} and have the effect of removing the (quasi) singularities.

Suppose the evolution operator is a combination of two evolution operators
\begin{equation}\label{U1U2}
   U(\calE)\PE=U_2(\calE)(\PEP+Q)U_1(\calE)
\end{equation}
with a possible intermediate model-space state of energy $\calE'$, which could lead to a (quasi)singularity. We can express this as
\begin{equation}\label{U1U2b}
   U(\calE)\PE=U_2(\calE)Q U_1(\calE)\PE+U_2(\calE)\GP(\calE') W_1(\calE)\PE,
\end{equation}
where
\begin{equation}\label{GEP}
   \GP(\calE')=\frac{\PEP}{\calE-\calE'}, \qquad \GEP W_1=U_1.
\end{equation}
 Then there will be a counter term
\begin{equation}\label{Counter}
   -U_2(\calE')\GEP W_1(\calE)\PE,
\end{equation}
where we note the energy parameter $\calE'$ in $U_2$. This leads to a "\bfit{model-space contribution}"
\begin{equation}\label{MSC}
   \pd{U_2}\PEP W_1\PE
\end{equation}
and the Green's operator, corresponding to the evolution operator \eqref{U1U2}
\begin{equation}\label{Green}
   \calG(\calE)=U_2(\calE)\GQ W_1(\calE)\PE+ \pd{U_2}\PEP W_1\PE.
\end{equation}
Here, the model-space contribution has eliminated the singularity due to the \ul{intermediate} model-space state. In addition, there might be other singularities, which can be handled similarly.

\vvsp
The operator~\cite[Ch. 6]{ILBook11}
\begin{equation}\label{R}
  {\cal R}=\Big(\im\partder{\calG(t,-\infty)}{t}\Big)_{t=0}
\end{equation}
is known as the \it{reaction operator} and closely related to the \it{effective interaction}
\begin{equation}\label{W}
  W=P {\cal R}P=P\Big(\im\partder{\calG(t,-\infty)}{t}\Big)_{t=0}P,
\end{equation}
and the \it{effective Hamiltonian} 
  \[H\eff=W+PH_0P\,;\qquad (H=H_0+V).\]
  The Green's operator has the same time dependence as the evolution operator\eqref{Ladd2}, and 
  applying the formula \eqref{R} to a Green's operator then eliminates the denominator of the resolvent \eqref{Resol}. Therefore, the Green's operator \eqref{Green} yields
  \begin{equation}\label{WGreen}
   W=PW_2(\calE)\GQ W_1(\calE)\PE+ P\pd{W_2}\PEP W_1\PE.
\end{equation}
This is the corresponding effective interaction, and the last term is the model-space contribution (MSC).

The Green's operator at finite final times for a single-particle system can generally be written in the form
\begin{eqnarray}\label{Gfin}
  &&\calG(t,-\infty)=\int\dif^3\bx\int\dif^4 x_1\,\hpsi\dagg(x)\SF(x,x_1){\cal F}\hpsi(x_1)\nn
  &=&c\dagg_m\bigbra{m}\intd{\om}\,\SF(\om,\bx,\bx_1){\cal F}\bigket{n}\,c_n\me^{-\im t(\om-\eps_m)}
  \me^{-\im t_1(\eps_n-\om)},
 \end{eqnarray}
 where $c_n\;c\dagg_m$ are electron creation/annihilation operators. We shall demonstrate below that the operator ${\cal F}$ is identical to the reaction operator ${\calR}$\eqref{R}.
 
Integration over $t_1$ and $\om$ gives $\om=\eps_n$
\begin{eqnarray*}  
  \calG(t,-\infty)=\me^{-\im t(\eps_n-\eps_m)}c\dagg_m\bra{m}\SF(\eps_n){\cal F}\ket{n}c_n
  =\me^{-\im t(\eps_n-H_0)}\SF(\eps_n){\cal F}.
\end{eqnarray*}
If we operate on a space $\PE$ with energy $\calE$, then $\eps_n=\calE$
\begin{eqnarray*}  
  \calG(t,-\infty)\PE=\me^{-\im t(\calE-H_0)}\SF(\calE)\cal F\PE.
\end{eqnarray*} 
If we here take the derivative with respect to $t$ at $t=0$, using \eqref{Resol}, we find from the definition \eqref{R} that the unknown operator ${\cal F}$ in \eqref{Gfin} is the reaction operator ${\calR}$.

If we let $t\rarr\infty$, then 
\begin{eqnarray}  
\im\calG(\infty,-\infty)&=&\int\dif^4 x\,\hpsi\dagg(x_1){\calR}(\bx_1)\hpsi(x_1)\,\me^{-\ga|t_1|}\nn
  &=&\int\dif t_1\,c\dagg_m\bra{m}{\calR}\ket{n}c_n\me^{-\im t_1(\eps_n-\eps_m)}\,\me^{-\ga|t_1|}.
\end{eqnarray}
Integration over $t_1$ gives 
\footnote{
\begin{equation} 
  \Delta_\ga(a)=\frac{1}{2\pi}\int_{-\infty}^\infty\dif t\,\me^{-ax}\me^{-\ga|t|}=\frac{1}{\pi}\frac{\ga}{a^2+\ga^2}
  \rarr\delta(a) \quad \rm{as}\quad\ga\rarr0
\end{equation}}
\begin{equation}  
  \im\calG(\infty,-\infty)P=2\pi\Delta_\ga(E_\rm{in}-E_\rm{out})\calR P\rarr2\pi\delta(E_\rm{in}-E_\rm{out})\calR P
\end{equation}
and the effective interaction
\begin{equation}  \label{GHeff}
   \boxed{P\im\calG(\infty,-\infty)P=2\pi\delta(E_\rm{in}-E_\rm{out})W.}
\end{equation}
If there are no intermediate model-space states, the Green's operator $\calG(\infty,-\infty)$ is identical to the S-matrix, which leads to the corresponding relation
\begin{equation}  \label{SHeff}
   P\im SP=2\pi\delta(E_\rm{in}-E_\rm{out})W.
\end{equation}

\bibliographystyle{prsty}

\input{RadRecombPubl5.bbl}

\end{document}

%% file: Commands.tex
\newcommand{\setl[1]}{\setlength{\unitlength}{#1cm}}
\newcommand{\nn}{\nonumber \\}
\newcommand{\eqqref}[1]{\mbox{Eq. (\ref{#1})}}
\newcommand{\eqpref}[1]{\mbox{(Eq. \ref{#1})}}
\newcommand{\eps}{\ensuremath{\varepsilon}}
\newcommand{\me}{\mathrm{e}}
\newcommand{\im}{\ensuremath{\mathrm{i}}}
\newcommand{\bra}[1]{\langle #1 |}
\newcommand{\ket}[1]{| #1 \rangle}
\newcommand{\proj}[1]{\ket{#1}\bra{#1}}
\newcommand{\Bra}[1]{\left\langle #1 \left|}
\newcommand{\Ket}[1]{\right| #1 \right\rangle}
\newcommand{\bigbra}[1] {\big\langle #1\big|}
\newcommand{\bigket}[1] {\big|#1\big\rangle}
\newcommand{\Bigbra}[1] {\Big\langle #1\Big|}
\newcommand{\Biggbra}[1] {\Bigg\langle #1\Bigg|}
\newcommand{\Bigket}[1] {\Big|#1\Big\rangle}
\newcommand{\Biggket}[1] {\Bigg|#1\Bigg\rangle}
\newcommand{\dif}{\ensuremath{\mathrm{d}}}
\newcommand{\difbx}{\dif^3\bx}
\newcommand{\difbk}{\dif^3\bk}
\newcommand{\difbp}{\dif^3\bp}
\newcommand{\intDim}[1]{\int\frac{\dif^D #1}{(2\pi)^D}}
\newcommand{\intD}[1]{\int\frac{\dif^3 #1}{(2\pi)^3}}
\newcommand{\intDD}[1]{\int\frac{\dif^4 #1}{(2\pi)^4}}
\newcommand{\intdinf}[1]{\int_{-\infty}^\infty\frac{\dif #1}{2\pi}}
\newcommand{\intbx}{\int\dif^3\bx}
\newcommand{\bx}{\boldsymbol{x}}
\newcommand{\bxdot}{\dot{\bx}}
\newcommand{\br}{\boldsymbol{r}}
\newcommand{\bk}{\boldsymbol{k}}
\newcommand{\bp}{\bs{\rm{p}}}
\newcommand{\bpi}{\boldsymbol{\pi}}
\newcommand{\bq}{\bs{\rm{q}}}
\newcommand{\bl}{\bs{\rm{l}}}
\newcommand{\bj}{\bs{\rm{j}}}
\newcommand{\bsp}{\bs{\rm{s}}}
\newcommand{\bL}{\bs{\rm{L}}}
\newcommand{\bJ}{\bs{\rm{J}}}
\newcommand{\bI}{\bs{\rm{I}}}
\newcommand{\bF}{\bs{\rm{F}}}
\newcommand{\bS}{\bs{\rm{S}}}
\newcommand{\hatl}{\hat{l}}
\newcommand{\hats}{\hat{s}}
\newcommand{\hatj}{\hat{j}}
\newcommand{\hatbj}{\hat{\bs{\rm{j}}}}
\newcommand{\hatbl}{\hat{\bs{\rm{l}}}}
\newcommand{\hatbs}{\hat{\bs{\rm{s}}}}
\newcommand{\hatL}{\hat{L}}
\newcommand{\hatS}{\hat{S}}
\newcommand{\hatJ}{\hat{J}}
\newcommand{\hatbJ}{\hat{\bs{\rm{J}}}}
\newcommand{\hatbL}{\hat{\bs{\rm{L}}}}
\newcommand{\hatbS}{\hat{\bs{\rm{S}}}}
\newcommand{\hatH}{\hat{H}}
\newcommand{\hrho}{\hat{\rho}}
\newcommand{\balpha}{\ensuremath{\boldsymbol{\alpha}}}
\newcommand{\bgamma}{\bs{\rm{\gamma}}}
\newcommand{\bgd}{\bs{\rm{\gamma}}\bsdot\!}
\newcommand{\bgp}{\bs{\rm{\gamma}}\bsdot\bp}
\newcommand{\bgq}{\bs{\rm{\gamma}}\bsdot\bq}
\newcommand{\bgk}{\bs{\rm{\gamma}}\bsdot\bk}
\newcommand{\gnp}{\gamma^0p_0}
\newcommand{\gnk}{\gamma^0k_0}
\newcommand{\giki}{\gamma^ik_i}
\newcommand{\gipi}{\gamma^ip_i}
\newcommand{\gmu}{\gamma^\mu}
\newcommand{\gnu}{\gamma^\nu}
\newcommand{\gs}{\gamma^\sigma}
\newcommand{\ga}{\gamma}
\newcommand{\wts}{\wt{\gamma}^\sigma}
\newcommand{\wtg}{\wt{\gamma}}
\newcommand{\gt}{\gamma^\tau}
\newcommand{\gb}{\gamma^\beta}
\newcommand{\gn}{\gamma^0}
\newcommand{\gi}{\gamma^i}
\newcommand{\gii}{\gamma_i}
\newcommand{\gj}{\gamma^j}
\newcommand{\gjj}{\gamma_j}
\newcommand{\gE}{\gamma_\rm{E}}
\newcommand{\dd}[1]{\frac{\partial}{\partial #1}}
\newcommand{\Partder}[1]{\frac{\partial }{\partial #1}}
\newcommand{\Partdern}[2]{\frac{\partial^#1 }{\partial#2 ^#1}}
\newcommand{\partder}[2]{\frac{\partial #1}{\partial #2}}
\newcommand{\partdern}[3]{{\frac{\partial^#1 #2}{\partial #3^#1}}}
\newcommand{\Cov}{\mathrm{Cov}}
\newcommand{\eff}{_{\mathrm{eff}}}
\newcommand{\SEp}{_{\mathrm{SEp}}}
\newcommand{\Irred}{\mathrm{Irred}}
\newcommand{\Sep}{\mathrm{Sep}}
\newcommand{\Nonsep}{\mathrm{Nonsep}}
\newcommand{\Irr}{_{\mathrm{Irr}}}
\newcommand{\RL}{_{\mathrm{L}}}
\newcommand{\QL}{_{\mathrm{QL}}}
\newcommand{\sgn}{\mathrm{sgn}}
\newcommand{\etc}{\mathrm{etc}}
\newcommand{\out}{\mathrm{out}}
\newcommand{\iin}{\mathrm{in}}
\newcommand{\inter}{\mathrm{int}}
\renewcommand{\rm}{\mathrm}
\newcommand{\mI}{\mathrm{I}}
\newcommand{\mS}{\mathrm{S}}
\newcommand{\mD}{\mathrm{D}}
\newcommand{\mG}{\mathrm{G}}
\newcommand{\mH}{\mathrm{H}}
\newcommand{\ext}{\mathrm{ext}}
\newcommand{\conn}{\mathrm{conn}}
\newcommand{\Counter}{\mathrm{Counter}}
\newcommand{\Linked}{\mathrm{linked}}
\newcommand{\linked}{\mathrm{linked}}
\newcommand{\mB}{\mathrm{B}}
\newcommand{\mC}{\mathrm{C}}
\newcommand{\SE}{\mathrm{SE}}
\newcommand{\G}{\mathrm{G}}
\newcommand{\sr}{\mathrm{sr}}
\newcommand{\op}{\mathrm{op}}
\newcommand{\opL}{\mathrm{opL}}
\newcommand{\cl}{\mathrm{cl}}
\newcommand{\clC}{\mathrm{clC}}
\newcommand{\bs}{\boldsymbol}
\newcommand{\bdot}{\boldsymbol\cdot}
\newcommand{\bnabla}{\bs{\nabla}}
\newcommand{\Q}{\mathcal{\boldsymbol{Q}}}
\newcommand{\bOm}{\mathcal{\boldsymbol{\Om}}}
\renewcommand{\H}{H}
\newcommand{\A}{A}
\newcommand{\bH}{\bs{\H}}
\newcommand{\bPsi}{\bs{\Psi}}
\newcommand{\bC}{\bs{C}}
\newcommand{\bR}{\bs{R}}
\newcommand{\bV}{\bs{V}}
\newcommand{\U}{U}
\newcommand{\Uop}{U_\op}
\newcommand{\Ucl}{U_\cl}
\newcommand{\p}{\hat{p}}
\newcommand{\hbp}{\hat{\bs{p}}}
\newcommand{\calH}{{\mathcal{H}}}
\newcommand{\calO}{\mathcal{O}}
\newcommand{\hO}{\hat{\calO}}
\newcommand{\calG}{\mathcal{G}}
\newcommand{\calP}{\mathcal{P}}
\newcommand{\Psit}{\widetilde{\Psi}}
\newcommand{\calGop}{\mathcal{G}_\op}
\newcommand{\calGcl}{\mathcal{G}_\cl}
\newcommand{\calV}{\mathcal{V}}
\newcommand{\calW}{\mathcal{U}}
\newcommand{\calU}{\mathcal{U}}
\newcommand{\calVsf}{\mathcal{V_\rm{sp}}}
\newcommand{\calL}{\mathcal{L}}
\newcommand{\IPair}{I^\rm{Pair}}
\newcommand{\dagg}{^{\dag}}
\newcommand{\intd}[1]{\int\frac{\dif #1}{2\pi}}
\newcommand{\half}{{\displaystyle\frac{1}{2}}}
\newcommand{\halfi}{{\displaystyle\frac{\im}{2}}}
\newcommand{\dint}{\int\!\!\!\int}
\newcommand{\tint}{\int\!\!\!\int\!\!\!\int}
\newcommand{\sint}{\tint\!\!\!\tint}
\newcommand{\ddint}{\dint\!\!\!\dint}
\newcommand{\dintd}[2]{\dint\frac{\dif #1}{2\pi}\,\frac{\dif #2}{2\pi}}
\newcommand{\tintd}[3]{\dint\frac{\dif #1}{2\pi}\,\frac{\dif #2}{2\pi}\,\frac{\dif #3}{2\pi}}
\newcommand{\ddintd}[4]{\ddint\frac{\dif #1}{2\pi}\,\frac{\dif #2}{2\pi}\,
\frac{\dif #3}{2\pi}\,\frac{\dif #4}{2\pi}}
\newcommand{\dddintd}[6]{\sint\frac{\dif #1}{2\pi}\,\frac{\dif #2}{2\pi}\,
\frac{\dif #3}{2\pi}\,\frac{\dif #4}{2\pi}\,\frac{\dif
#5}{2\pi}\,\frac{\dif #6}{2\pi}}
\newcommand{\ddt}{\frac{\partial}{\partial t}}
\newcommand{\intbr}{\int\dif\br\,}
\newcommand{\gamlim}{\ensuremath{\gamma\rightarrow 0}}
\newcommand{\vsp}{\vspace{0.5cm}}
\newcommand{\vvsp}{\vspace{0.25cm}}
\newcommand{\vvvsp}{\vspace{0.125cm}}
\newcommand{\mvsp}{\vspace{-0.5cm}}
\newcommand{\mvvsp}{\vspace{-0.25cm}}
\newcommand{\mvvvsp}{\vspace{-0.25cm}}
\newcommand{\mmvsp}{\vspace{-0.25cm}}
\newcommand{\mVsp}{\vspace{-1cm}}
\newcommand{\mVSP}{\vspace{-2cm}}
\newcommand{\Vsp}{\vspace{1cm}}
\newcommand{\Wsp}{\vspace{2cm}}
\newcommand{\hsp}{\hspace{0.5cm}}
\newcommand{\hhsp}{\hspace{0.25cm}}
\newcommand{\Hsp}{\hspace{1cm}}
\newcommand{\HSP}{\hspace{2cm}}
\newcommand{\mmhsp}{\hspace{-0.25cm}}
\newcommand{\mhsp}{\hspace{-0.5cm}}
\newcommand{\mHsp}{\hspace{-1cm}}
\newcommand{\mHSP}{\hspace{-2cm}}
\newcommand{\ö}{\"{o}}
\newcommand{\š}{\"{o}}
\newcommand{\Š}{\"{a}}
\newcommand{\Ö}{\"{O}}
\newcommand{\…}{\"{O}}
\newcommand{\ä}{\"a}
\newcommand{\å}{\aa}
\newcommand{\Å}{\AA}
\newcommand{\ue}{\"{u}}
\newcommand{\hpsi}{\hat{\psi}}
\newcommand{\wt}[1]{\widetilde{#1}}
\newcommand{\WU}{\widetilde{U}}
\renewcommand{\it}{\textit}
\renewcommand{\bf}{\textbf}
\newcommand{\bfit}[1]{\textbf{\it{#1}}}
\newcommand{\bful}[1]{\textbf{\ul{#1}}}
\newcommand{\ul}{\underline}
\newcommand{\itul}[1]{\it{\ul{#1}}}
\newcommand{\abs}[1]{|{#1}|}
\newcommand{\Abs}[1]{\big|{#1}\big|}
\newcommand{\eq}{\eqref}
\newcommand{\rarr}{\rightarrow}
\newcommand{\larr}{\leftarrow}
\newcommand{\lrarr}{\leftrightarrow}
\newcommand{\LRarr}{\Longleftrightarrow}
\newcommand{\Rarr}{\Rightarrow}
\newcommand{\Lrarr}{\Longrightarrow}
\newcommand{\rr}{r_{12}}
\newcommand{\frr}{\frac{1}{r_{12}}}
\newcommand{\brr}{\br_{12}}
\newcommand{\DF}{D_{\rm{F}\nu\mu}}
\newcommand{\DFS}{D_{\rm{F}}}
\newcommand{\DFmn}{D_{\rm{F}\mu\nu}}
\newcommand{\DFCij}{D^C_{\rm{Fij}}}
\newcommand{\SF}{S_{\rm{F}}}
\newcommand{\hSF}{\hat{S}_{\rm{F}}}
\newcommand{\hh}{\hat{h}}
\newcommand{\QED}{\rm{QED}}
\newcommand{\CQED}{\rm{CQED}}
\newcommand{\DQED}{\rm{DQED}}
\newcommand{\ren}{\rm{ren}}
\newcommand{\bou}{\rm{bou}}
\newcommand{\free}{\rm{free}}
\renewcommand{\sp}[2]{\bra{#1}{#2}\rangle}
\newcommand{\SP}[2]{\Bigbra{#1}{#2}\Big\rangle}
\newcommand{\qand}{\quad\rm{and}\quad}
\newcommand{\V}{V}
\newcommand{\VC}{V_\rm{C}}
\newcommand{\VF}{V_\rm{F}}
\newcommand{\VG}{V_\rm{G}}
\newcommand{\VspC}{V_\rm{TC}}
\newcommand{\Vsr}{V_\rm{sr}}
\newcommand{\MC}{\calM_\rm{C}}
\newcommand{\IC}{I^\rm{C}}
\newcommand{\ICC}{I^\rm{C}_\rm{C}}
\newcommand{\ICT}{I^\rm{C}_\rm{T}}
\newcommand{\UT}{U_\rm{T}}
\newcommand{\VT}{V_\rm{T}}
\newcommand{\vT}{v_\rm{T}}
\newcommand{\HD}{H_\rm{D}}
\newcommand{\fC}{f^\rm{C}}
\newcommand{\fCC}{f^\rm{C}_\rm{C}}
\newcommand{\fCT}{f^\rm{C}_\rm{T}}
\newcommand{\fF}{f^\rm{F}}
\newcommand{\VB}{V_\rm{B}}
\newcommand{\Vsf}{V_\rm{sp}}
\newcommand{\Vgsf}{\calV_\rm{T}}
\newcommand{\Vsfp}{V_\rm{sp}}
\newcommand{\Usfp}{U_\rm{sp}}
\newcommand{\Msfp}{\calM_\rm{sp}}
\newcommand{\MVx}{\calM_\rm{Vx}}
\newcommand{\VSE}{V_\rm{SE}}
\newcommand{\VVert}{V_\rm{Vx}}
\newcommand{\VVx}{V_\rm{Vx}}
\newcommand{\VspGen}{V_\rm{tr}^\rm{Gen}}
\newcommand{\VTC}{V_\rm{TC}}
\newcommand{\calUsp}{\calU_\rm{sp}}
\newcommand{\Usf}{U_\rm{sp}}
\newcommand{\Ssf}{S_\rm{T}}
\newcommand{\Msf}{\calM_\rm{sp}}
\newcommand{\MT}{\calM_\rm{T}}
\newcommand{\Ksf}{{\cal K_\rm{sp}}}
\newcommand{\MSE}{\calM_\rm{SE}}
\newcommand{\SSE}{S_\rm{SE}}
\newcommand{\USE}{U_\rm{SE}}
\newcommand{\Coul}{\rm{Coul}}
\newcommand{\calK}{\mathcal{\kappa}}
\newcommand{\calKc}{I_c}
\newcommand{\calF}{\mathcal{F}}
\newcommand{\calM}{{\mathcal{M}}}
\newcommand{\calR}{\mathcal{\hat{R}}}
\newcommand{\calB}{\mathcal{B}}
\newcommand{\GQ}{\Gamma_Q}
\newcommand{\GV}{\Gamma V}
\newcommand{\Gam}{\Gamma}
\newcommand{\GQV}{\GQ V}
\newcommand{\GI}{\calG^\rm{I}}
\newcommand{\bGQ}{\bs{\Gamma_Q}}
\newcommand{\Gk}{\bs{\Gamma_Q}}
\newcommand{\Gv}{\Gamma_Q}
\newcommand{\TD}{T_\rm{D}}
\newcommand{\F}{\rm{F}}
\renewcommand{\P}{\bs{P}}
\newcommand{\Util}{\widetilde{U}}
\newcommand{\Ucov}{U_\Cov}
\renewcommand{\S}{S}
\newcommand{\Vtil}{\widetilde{V}}
\newcommand{\Om}{\Omega}
\newcommand{\Omsf}{\Omega_\rm{sp}}
\newcommand{\Omi}{\Om_\rm{I}}
\newcommand{\OmI}{{\Om_\rm{I}}}
\newcommand{\om}{\omega}
\newcommand{\OM}{\mathcal{\boldsymbol{\Om}}}
\newcommand{\Ombar}{\bar{\Omega}}
\newcommand{\Vbar}{\bar{V}}
\newcommand{\VI}{V_{12}}
\newcommand{\VR}{V_{\rm{R}}}
\newcommand{\VRbar}{\Bar{V}_{\rm{R}}}
\newcommand{\calE}{{\mathcal{E}}}
\newcommand{\CALE}{{\mathcal{E}}}
\newcommand{\la}{\lambda}
\newcommand{\FLL}{F_\rm{LL}}
\newcommand{\ka}{|\bs{k}|}
\newcommand{\norm}[1]{||{#1}||}
\newcommand{\Norm}[1]{\big|\big|{#1}\big|\big|}
\newcommand{\NORM}[1]{\Big|\Big|{#1}\Big|\Big|}
\newcommand{\Fr}{Fr\'echet }
\newcommand{\Ga}{G\^ateaux }
\newcommand{\epsn}{\ensuremath{\epsilon}}
\newcommand{\Der}[1]{\frac{\dif}{\dif#1}}
\newcommand{\der}[2]{\frac{\dif#1}{\dif#2}}
\newcommand{\ave}[1]{\langle #1\rangle}
\newcommand{\Ave}[1]{\Big\langle #1\Big\rangle}
\newcommand{\BB}[2]{\Big\{{#1}\,\Big|\:{#2}\Big\}}
\newcommand{\LL}{L^1\cap L^3}
\newcommand{\partdelta}[2]{\frac{\delta #1}{\delta #2}}
\newcommand{\pd}[1]{\frac{\delta #1}{\delta \calE}}
\newcommand{\pda}[1]{\frac{\delta^* #1}{\delta \calE}}
\newcommand{\Pda}[1]{\frac{\delta^*}{\delta #1}}
\newcommand{\Pdna}[2]{\frac{\delta^{^*#1}}{\delta #2}}
\newcommand{\Partdelta}[1]{\frac{\delta }{\delta #1}}
\newcommand{\Pd}[1]{\frac{\delta }{\delta #1}}
\newcommand{\partdeltan}[3]{\frac{\delta^#1 #2}{\delta #3^#1}}
\newcommand{\partdeltanp}[3]{\frac{\delta^{(#1)} #2}{\delta
#3^{(#1)}}}
\newcommand{\pdna}[2]{\frac{\delta^{^*#1} #2}{\delta\calE^#1}}
\newcommand{\pdn}[2]{\frac{\delta^{#1} #2}{\delta\calE^{#1}}}
\newcommand{\pdnp}[2]{\frac{\delta^{(#1)} #2}{\delta\calE^{(#1)}}}
 \newcommand{\ett}{^{(1)}}
\newcommand{\nol}{^{(0)}}
\newcommand{\tva}{^{(2)}}
\newcommand{\tre}{^{(3)}}
\newcommand{\fyr}{^{(4)}}
\newcommand{\enh}{^{(1/2)}}
\newcommand{\treh}{^{(3/2)}}
\newcommand{\femh}{^{(5/2)}}
\newcommand{\n}{^{(n)}}
\newcommand{\m}{^{(m)}}
\newcommand{\nm}{^{(n-m)}}
\newcommand{\nett}{^{(n-1)}}
\newcommand{\ntva}{^{(n-2)}}
\newcommand{\PE}{P_\mathcal{E}}
\newcommand{\GE}{\Gamma(\calE)}
\newcommand{\GEE}{\Gamma(E)}
\newcommand{\GQEE}{\Gamma_Q(E)}
\newcommand{\GQE}{\Gamma_Q(\calE)}
\newcommand{\GQEP}{\Gamma_Q(\EP)}
\newcommand{\GEN}{\Gamma(E_0)}
\newcommand{\GQEN}{\Gamma_Q(E_0)}
\newcommand{\GEP}{\Gamma(\calE')}
\newcommand{\GaV}{\Gamma V}
\newcommand{\GaVg}{\GQ V}
\newcommand{\PEP}{P_{\calE'}}
\newcommand{\QE}{Q_{\calE}}
\newcommand{\QEP}{Q_{\calE'}}
\newcommand{\PEPP}{P_{\calE''}}
\newcommand{\EP}{\calE'}
 \newcommand{\limgam}{\lim_{\gamlim}}
 \newcommand{\Ugam}[1]{U_\gamma(#1,-\infty)}
\newcommand{\Ugamtil}[1]{\widetilde{U}_\gamma(#1,-\infty)}
\newcommand{\Ugamt}{\widetilde{U}_\gamma}
\newcommand{\img}{\im\gamma}
\newcommand{\ime}{\im\eta}
\newcommand{\pbar}{\not\!p}
\newcommand{\psl}{\!\not\!p\,}
\newcommand{\Asl}{\not\!\! A\,}
\newcommand{\qsl}{\not\!q}
\newcommand{\ksl}{\not\!k}
\newcommand{\lsl}{\not\!l}
\newcommand{\nott}{\not\!\!}
\newcommand{\wtp}{\widetilde{p}}
\newcommand{\wtq}{\widetilde{q}}
\newcommand{\wtk}{\widetilde{k}}
\newcommand{\Deltag}{\Delta_\gamma}
\newcommand{\Deltatg}{\Delta_{2\gamma}}
\newcommand{\Deltafg}{\Delta_{4\gamma}}
\renewcommand{\bar}{\setlength{\unitlength}{0.6cm}\put(0,0.6){\line(1,0){0.4}}}
\newcommand{\fbar}{\setlength{\unitlength}{0.6cm}\put(0,0.4){\line(1,0){0.4}}}
\newcommand{\MSC}{\rm{MSC}}
\newcommand{\IF}{\rm{IF}}
                                                  \newcommand{\Htil}{\widetilde{H}}
                                                  \newcommand{\Ubar}{\bar{U}}
                                                  \newcommand{\Hbar}{\bar{V}}
                                                  \newcommand{\Udot}{\dot{U}}
                                                  \newcommand{\Ubardot}{\dot{\Ubar}}
                                                  \newcommand{\Utildot}{\dot{\Util}}
                                                  \newcommand{\Cdot}{\dot{C}}
                                                  \newcommand{\Ombardot}{\dot{\Ombar}}
                                                  \newcommand{\bsdot}{\bs{\cdot}}

\newcommand{\npartdelta}[3]{\frac{\delta^#1 #2}{\delta #3^#1}}
\newcommand{\ip}[1]{| #1 \rangle \langle #1 |}
\newcommand{\con}{\mathrm{con}}
\newcommand{\ph}{\mathrm{ph}}
\newcommand{\bA}{\bs{A}}
\newcommand{\Pmu}{\partial^\mu}
\newcommand{\Pnu}{\partial^\nu}
\newcommand{\pmu}{\partial_\mu}
\newcommand{\pnu}{\partial_\nu}
\newcommand{\bAT}{\bs{A}_\perp}
\newcommand{\bAL}{\bs{A}_\parallel}
\newcommand{\bET}{\bs{E}_\perp}
\newcommand{\bEL}{\bs{E}_\parallel}
\newcommand{\bB}{\bs{\rm{B}}}
\newcommand{\bE}{\bs{\rm{E}}}
\newcommand{\bn}{\bs{n}}
\newcommand{\beps}{\bs{\eps}}
\newcommand{\HI}{H_\rm{int,I}}
\newcommand{\calHI}{{\cal H}_\rm{int,I}}
\newcommand{\ih}{\frac{\im}{\hbar}}
\newcommand{\bkx}{\bk\bdot\bx}
\newcommand{\halfS}{{\textstyle\frac{1}{2}\,}}
\newcommand{\h}{\hat{h}}
\newcommand{\DFnu}{D_{\rm{F}\nu\mu}}
\newcommand{\Eab}{E_{ab}}
\newcommand{\rE}{{\red E}}
\newcommand{\sphline}{\hline\vsp}
\newcommand{\Ram}[4]
{\begin{picture}(0,0)(0,0)\setlength{\unitlength}{1cm}
    \put(#1,#2){\Ebox{#3}{#4}}
  \end{picture}}
                                                  \newcommand{\Ers}{E_{rs}}
                                                  \newcommand{\Etu}{E_{tu}}
                                                  \newcommand{\Eru}{E_{ru}}
                                                  \newcommand{\Ecd}{E_{cd}}
                                                  \newcommand{\Erd}{E_{rd}}
                                                  \newcommand{\epsi}{\epsilon}

%% file: FigurecommandsV.tex

\newcommand{\NVP}
{\put(0,0){\LineV{4}}\put(2,0){\LineV{4}}}
\newcommand{\SVP}
{\put(0,0){\LineV{2.5}}\put(-1,1.5){\LineV{2.5}}\put(0,2.5){\LineDl{}}
\put(2,0){\LineV{4}}}
\newcommand{\DVP}
{\put(0,0){\LineV{2.5}}\put(-1,1.5){\LineV{2.5}}\put(0,2.5){\LineDl{}}
\put(2,0){\LineV{2.5}}\put(3,1.5){\LineV{2.5}}\put(2,2.5){\LineDr{}}}
\newcommand{\SVPC}
{\put(0,0){\LineV{3}}\put(-1,2){\LineV{2}}\put(0,3){\LineDl{}}
\put(2,0){\LineV{4}}}
\newcommand{\DVPC}
{\put(0,0){\LineV{3}}\put(-1,2){\LineV{2}}\put(0,3){\LineDl{}}
\put(2,1){\LineV{3}}\put(3,0){\LineV{2}}\put(3,2){\LineDl{}}}

\newcommand{\circl}[0]
{\circle*{0.225} }

\newcommand{\LineH}[1]
{\linethickness{0.5mm} \put(0,0){\line(1,0){#1}} }

\newcommand{\LineHH}[1]
{\linethickness{1mm} \put(0,0){\line(1,0){#1}} }

\newcommand{\LineD}[1]
{\put(0,-0.2){\line(1,0){#1}} \put(0,0.2){\line(1,0){#1}}}

\newcommand{\LineDD}[1]
{\put(0,-0.3){\line(1,0){#1}} \put(0,0.3){\line(1,0){#1}}}

\newcommand{\LineS}[1]
{\linethickness{1mm}
\put(0,0){\line(1,0){#1}}}

\newcommand{\LineWO}[1]
{\linethickness{0.75mm} \put(0,0){\line(1,0){#1}} }

\newcommand{\LineV}[1]
{\linethickness{0.5mm}  \put(0,0){\line(0,1){#1}} }

\newcommand{\lineD}[1]
{\linethickness{0.1mm}  \put(-0.035,0){\line(0,1){#1}}\put(0.035,0){\line(0,1){#1}} }

\newcommand{\LineEtt}[2]
{\linethickness{0.5mm}  \put(0,0){\line(0,1){#1}}
\put(0,#2){\VectorUp}}

\newcommand{\LineTva}[3]
{\linethickness{0.5mm}  \put(0,0){\line(0,1){#1}}
\put(0,#2){\VectorUp}\put(0,#3){\VectorUp} }

\newcommand{\LineTre}[4]
{\linethickness{0.5mm}  \put(0,0){\line(0,1){#1}}
\put(0,#2){\VectorUp}\put(0,#3){\VectorUp}\put(0,#4) {\VectorUp}}

\newcommand{\LineFyr}[5]
{\linethickness{0.5mm}  \put(0,0){\line(0,1){#1}}
\put(0,#2){\VectorUp}\put(0,#3){\VectorUp}\put(0,#4){\VectorUp}\put(0,#5)
{\VectorUp} }

\newcommand{\LineFem}[6]
{\linethickness{0.5mm}  \put(0,0){\line(0,1){#1}}
\put(0,#2){\VectorUp}\put(0,#3){\VectorUp}\put(0,#4){\VectorUp}\put(0,#5)
{\VectorUp}\put(0,#6){\VectorUp} }

\newcommand{\LineSex}[7]
{\linethickness{0.5mm}  \put(0,0){\line(0,1){#1}} 
\put(0,#2){\VectorUp}\put(0,#3){\VectorUp}\put(0,#4){\VectorUp}\put(0,#5)
{\VectorUp}\put(0,#6){\VectorUp}\put(0,#7){\VectorUp}
}

\newcommand{\LineVT}[1]
{\linethickness{0.3mm}  \put(0,0){\line(0,1){#1}} }

\newcommand{\Linev}[1]
{\put(0,0){\line(0,1){#1}} }

\newcommand{\LineW}[1]
{\linethickness{0.75mm} \put(0,0){\line(0,1){#1}}}

\newcommand{\LineHpt}[1]
{\put(0,0.015){\line(1,0){#1}} \put(0,-0.015){\line(1,0){#1}}
\put(0,0){\circl} \put(#1,0){\circl}}

\newcommand{\LineDl}[1]
{\put(0.012,-0.012){\line(-1,-1){#1}}\put(-0.012,0.012){\line(-1,-1){#1}}
\put(0.0,0.0){\line(-1,-1){#1}}}

\newcommand{\LineUl}[1]
{\put(0.012,-0.012){\line(-1,1){#1}}\put(-0.012,0.012){\line(-1,1){#1}}
\put(0.0,0.0){\line(-1,1){#1}}}

\newcommand{\Lineul}[1]
{\put(0.012,-0.012){\line(-1,2){#1}}\put(-0.012,0.012){\line(-1,2){#1}}
\put(0.0,0.0){\line(-1,2){#1}}}

\newcommand{\Linedl}[1]
{\put(0.01,0.01){\line(-1,-2){#1}}
\put(-0.01,-0.01){\line(-1,-2){#1}}}

\newcommand{\LineUr}[1]
{\put(0,0.015){\line(1,1){#1}}
\put(0,0){\line(1,1){#1}}
\put(0,-0.015){\line(1,1){#1}}}

\newcommand{\LineDr}[1]
{\put(0,0.01){\line(1,-1){#1}} \put(0,0){\line(1,-1){#1}}
\put(0,-0.01){\line(1,-1){#1}}}

\newcommand{\Linedr}[1]
{\put(0,0.01){\line(1,-2){#1}} \put(0,0){\line(1,-2){#1}}
\put(0,-0.01){\line(1,-2){#1}}}

\newcommand{\Lineur}[1]
{\put(0,0.025){\line(1,2){#1}} \put(0,0){\line(1,2){#1}}
\put(0,-0.025){\line(1,2){#1}}}

\newcommand{\DLine}[1]
{\put(0,-0.05){\line(1,0){#1}} \put(0,0.05){\line(1,0){#1}}
\put(0,0){\circl}}

\newcommand{\Vector}[0]
{\thicklines\put(-0.10,-0.035){\vector(-1,0){0}}}

\newcommand{\VectorR}[0]
{\thicklines\put(0.14,0.035){\vector(1,0){0}}}

\newcommand{\vectorR}[0]
{\put(0.13,0.1){\vector(1,0){0}}}

\newcommand{\VectorUp}[0]
{\thicklines
\put(0,0.15){\vector(0,0){0}}}

\newcommand{\vectorUp}[0]
{\setlength{\unitlength}{1cm} \put(0,0.12){\vector(0,0){0}} }

\newcommand{\VectorT}[0]
{\thicklines\setlength{\unitlength}{1cm}\put(0,0.18){\vector(0,0){0}}
\put(0.02,0.02){\vector(0,0){0}}}

\newcommand{\VectorDn}[0]
{\thicklines\setlength{\unitlength}{1cm}
\put(0,-0.15){\vector(0,-1){0}}}

\newcommand{\VectorDl}[0]
{\thicklines \setlength{\unitlength}{1cm}
\put(-0.1,-0.1){\vector(-1,-1){0}}}

\newcommand{\VectorDr}[0]
{\thicklines\setlength{\unitlength}{1cm}
\put(0.084,-0.092){\vector(1,-1){0}}}

\newcommand{\Vectordr}[0]
{\thicklines\setlength{\unitlength}{1cm}
\put(0.022,0.112){\vector(1,-3){0}} }

\newcommand{\Vectorur}[0]
{\thicklines\setlength{\unitlength}{1cm}
\put(0.04,-0.062){\vector(1,2){0}} }

\newcommand{\VectorUr}[0]
{\thicklines\setlength{\unitlength}{1cm}
 \put(0.22,-0.1){\vector(1,1){0}}}

\newcommand{\VectorUl}[0]
{\put(-0.23,-0.02){\vector(-1,1){0}}
\put(-0.19,-0.03){\vector(-1,1){0}}
\put(-0.22,-0.06){\vector(-1,1){0}} }

\newcommand{\Vectorul}[0]
{\put(-0.13,-0.02){\vector(-1,2){0}}
\put(-0.11,-0.04){\vector(-1,2){0}}
\put(-0.16,-0.06){\vector(-1,2){0}}}

\newcommand{\Wector}[0]
{\put(-0.15,0){\Vector}\put(0.15,0){\Vector}}

\newcommand{\WectorUp}[0]
{\put(0,0.125)\VectorUp\put(0,-0.125)\VectorUp}

\newcommand{\WectorDn}[0]
{\put(0,0.125)\VectorDn\put(0,-0.125)\VectorDn}

\newcommand{\WectorDl}[0]
{\put(0.1,0.1)\VectorDl\put(-0.1,-0.1)\VectorDl}

\newcommand{\Wectordl}[0]
{\setlength{\unitlength}{1cm}
 \put(0.04,0.10){\vector(-2,-1){0}}\put(0.02,0.13){\vector(-2,-1){0}}
 \put(-0.14,0.02){\vector(-2,-1){0}}\put(-0.16,0.05){\vector(-2,-1){0}}}

\newcommand{\EllineH}[4]
{\put(0,0){\LineH{#1}} \put(#2,0){\Vector}
\put(#2,0.45){\makebox(0,0){$#3$}}
\put(#2,-0.35){\makebox(0,0){$#4$}}}

\newcommand{\lline}[4]
{\put(0,0){\LineV{#1}} \put(-0.3,#2){\makebox(0,0){$#3$}}
\put(0.4,#2){\makebox(0,0){$#4$}}}

\newcommand{\Elline}[4]
{\put(0,0){\LineV{#1}} \put(0,#2){\VectorUp}
\put(-0.35,#2){\makebox(0,0){$#3$}}
\put(0.4,#2){\makebox(0,0){$#4$}} }

\newcommand{\EllineDV}[3]
{\put(0,0){\LineV{#1}}\put(0,#2){\VectorUp} \put(0,#3){\VectorUp}}

\newcommand{\EllineTV}[4]
{\put(0,0){\LineV{#1}}\put(0,#2){\VectorUp}
\put(0,#3){\VectorUp}\put(0,#4){\VectorUp}}

\newcommand{\EllineFV}[5]
{\put(0,0){\LineV{#1}}\put(0,#2){\VectorUp}
\put(0,#3){\VectorUp}\put(0,#4){\VectorUp}\put(0,#5){\VectorUp}}

\newcommand{\EllineSV}[6]
{\put(0,0){\LineV{#1}}\put(0,#2){\VectorUp}
\put(0,#3){\VectorUp}\put(0,#4){\VectorUp}\put(0,#5){\VectorUp}
\put(0,#6){\VectorUp}}

\newcommand{\elline}[3]
{\put(0,0){\LineV{#1}} \put(0,0){\makebox(0,#1){\VectorUp}}
\put(-0.4,0){\makebox(0,#1){$#2$}}
\put(0.5,0){\makebox(0,#1){$#3$}}
}

\newcommand{\ellineP}[6]
{\put(0,0){\elline{#1}{#3}{#4}}\put(#2,0){\elline{#1}{#5}{#6}} 
}

\newcommand{\EllineW}[4]
{\put(0,0){\LineW{#1}} \put(0,#2){\WectorUp}
\put(-0.4,#2){\makebox(0,0){$#3$}}
\put(0.5,#2){\makebox(0,0){$#4$}} }

\newcommand{\Ellinev}[4]
{\put(0,0){\Linev{#1}} \put(0,#2){\VectorUp}
\put(-0.3,#2){\makebox(0,0){$#3$}}
\put(0.4,#2){\makebox(0,0){$#4$}} }

\newcommand{\DElline}[4]
{\put(0,0){\LineV{#1}} \put(0,#2){\WectorUp}
\put(-0.3,#2){\makebox(0,0){$#3$}}
\put(0.3,#2){\makebox(0,0){$#4$}}}

\newcommand{\DEllineDn}[4]
{\put(0,0){\LineV{#1}} \put(0,#2){\WectorDn}
\put(-0.25,#2){\makebox(0,0){$#3$}}
\put(0.25,#2){\makebox(0,0){$#4$}}}

\newcommand{\Ellinet}[4]
{\put(0,0){\Linev{#1}} \put(0,#2){\vector(0,1){0}}
\put(-0.35,#2){\makebox(0,0){$#3$}}
\put(0.35,#2){\makebox(0,0){$#4$}}}

\newcommand{\EllineT}[4]
{\put(0,0){\LineW{#1}} \put(0,#2){\VectorUp}
\put(-0.35,#2){\makebox(0,0){$#3$}}
\put(0.35,#2){\makebox(0,0){$#4$}}}

\newcommand{\EllineDnt}[4]
{\put(0,0){\Linev{#1}} \put(0,#2){\VectorDn}
\put(-0.35,#2){\makebox(0,0){$#3$}}
\put(0.35,#2){\makebox(0,0){$#4$}}}

\newcommand{\EllineDn}[4]
{\put(0,0){\LineV{#1}} \put(0,#2){\VectorDn}
\put(-0.35,#2){\makebox(0,0){$#3$}}
\put(0.5,#2){\makebox(0,0){$#4$}}}

\newcommand{\EllineDl}[4]
{\put(0,0){\LineDl{#1}} \put(-#2,-#2){\VectorDl}
\put(-#2,-#2){\makebox(-0.25,0.25){$#3$}}
\put(-#2,-#2){\makebox(0.5,-0.25){$#4$}}}

\newcommand{\Ellinedl}[4]
{\put(0.01,0.01){\line(-1,-2){#1}}
\put(-0.01,-0.01){\line(-1,-2){#1}}
\thicklines\put(-0.05,-0.1){\vector(-1,-2){#2}}
\put(-0.4,-0.4){\makebox(-#1,-#1){$#3$}}
\put(-0.5,-0.6){\makebox(-#1,-#1){$#4$}}}

\newcommand{\EllinedL}[4]
{\put(0.02,0.02){\line(-1,-3){#1}}
\put(-0.02,-0.02){\line(-1,-3){#1}}
\thicklines\put(-0.05,-0.1){\vector(-1,-3){#2}}
\put(-0.4,0){\makebox(-1,-3){$#3$}}
\put(0.4,-0){\makebox(-1,-3){$#4$}}}

\newcommand{\Ellinedr}[4]
{\put(0.01,0.01){\line(1,-2){#1}}
\put(-0.01,-0.01){\line(1,-2){#1}}
\thicklines\put(0.05,-0.1){\vector(1,-2){#2}}
\put(-0.4,0){\makebox(1,-2){$#3$}}
\put(0.2,0.2){\makebox(1,-2){$#4$}}}

\newcommand{\EllineA}[7]
{\put(0.0,0.0){\line(#1,#2){#3}} \put(0.005,0.0){\line(#1,#2){#3}}
\put(-0.005,0.0){\line(#1,#2){#3}} \put(0,0){\vector(#1,#2){#4}}
\put(0.010,0){\vector(#1,#2){#4}}
\put(-0.010,0){\vector(#1,#2){#4}}
\put(#6,#7){\makebox(0,0){$#5$}}}

\newcommand{\EllineDr}[4]
{\put(0,0){\LineDr{#1}} \put(#2,-#2){\makebox(0.2,0.2)\VectorDr}
\put(#2,-#2){\makebox(-0.4,-0.3){$#3$}}
\put(#2,-#2){\makebox(0.75,0.25){$#4$}}}

\newcommand{\EllinedR}[5]
{\put(0,0){\line(1,-3){#1}} \put(0.014,0){\line(1,-3){#1}}
\put(-0.014,0){\line(1,-3){#1}}
\put(#2,-#3){\makebox(0,0){{\Vectordr}}}
\put(#2,-#3){\makebox(-0.5,-0.5){$#4$}}
\put(#2,-#3){\makebox(0.5,0.5){$#5$}}}

\newcommand{\EllineuR}[5]
{\put(0,0){\line(1,3){#1}} \put(0.014,0){\line(1,3){#1}}
\put(-0.014,0){\line(1,3){#1}}
\put(#2,#3){\makebox(0,0){\Vectorur}}
\put(#2,#3){\makebox(-0.5,-0.5){$#4$}}
\put(#2,#3){\makebox(0.5,0.5){$#5$}}}

\newcommand{\EllineUl}[4]
{\put(0,0){\LineUl{#1}} \put(-#2,#2){{\makebox(0,-0.15)\VectorUl}}
\put(-#2,#2){\makebox(-0.5,0.5){$#3$}}
\put(-#2,#2){\makebox(0.5,0.5){$#4$}}}

\newcommand{\Ellineul}[5]
{\put(0,0){\Lineul{#1}} \put(#2,#3){\makebox(0.5,0){\Vectorul}}
\put(#2,#3){\makebox(-0.5,0){$#4$}}\put(#2,#3){\makebox(0.5,0){$#5$}}
}

\newcommand{\Ellineur}[5]
{\put(0,0){\Lineur{#1}}
\put(#2,#3){\makebox(0,0){\Vectorur}}
\put(#2,#3){\makebox(-0.5,0){$#4$}}
\put(#2,#3){\makebox(0.5,0){$#5$}}}

\newcommand{\EllineUr}[4]
{\put(0,0){\LineUr{#1}} \put(#2,#2){\makebox(-0.35,0){\VectorUr}}
\put(-0.2,0.4){\makebox(#1,#1){$#3$}}
\put(0.2,-0.1){\makebox(#1,#1){$#4$}}}

\newcommand{\DEllineDl}[4]
{\put(0,0){\LineDl{#1}}
\put(-#2,-#2){\WectorDl}
\put(-0.25,0.25){\makebox(-#1,-#1){$#3$}}
\put(0.25,-0.25){\makebox(-#1,-#1){$#4$}}}

\newcommand{\Ebox}[2]
{\put(0,0){\LineH{#1}} \put(0,#2){\LineH{#1}}
\put(0,0){\LineV{#2}} \put(#1,0){\LineV{#2}}}

\newcommand{\EEbox}[2]
{\put(0,0){\LineHH{#1}} \put(0,#2){\LineHH{#1}}
\put(0,0){\LineW{#2}} \put(#1,0){\LineW{#2}}}

\newcommand{\dashH}
{\multiput(0.05,0)(0.25,0){5}{\line(1,0){0.15}}}

\newcommand{\dash}[1]
{\multiput(0.05,0)(0.25,0){#1}{\line(1,0){0.15}}}

\newcommand{\dashV}[1]
{\multiput(0.05,0)(0,0.25){#1}{\line(0,1){0.15}}}

\newcommand{\dashHp}
{\multiput(0.05,0)(0.25,0){6}{\line(1,0){0.15}}}

\newcommand{\DashH}
{\multiput(0.05,0)(0.25,0){10}{\line(1,0){0.15}}}

\newcommand{\dashHnum}[2]
{\multiput(0.05,0)(0.25,0){5}{\line(1,0){0.15}}
\put(-0.25,0){\makebox(0,0){$#1$}}
\put(1.5,0){\makebox(0,0){$#2$}}}

\newcommand{\dashHnuma}[2]
{\multiput(0.05,0)(0.25,0){5}{\line(1,0){0.15}}
\put(0.25,0.25){\makebox(0,0){$#1$}}
\put(1,0.25){\makebox(0,0){$#2$}}}

\newcommand{\dashHnumu}[2]
{\multiput(0.05,0)(0.25,0){5}{\line(1,0){0.15}}
\put(0.25,-0.25){\makebox(0,0){$#1$}}
\put(1,-0.25){\makebox(0,0){$#2$}}}

\newcommand{\Potint}
{\put(0,0)\dashH \put(1.35,0){\makebox(0,0){$\times$}}
\put(0,0){\circle*{0.15}} }

\newcommand{\potint}
{\multiput(0.05,0)(0.25,0){3}{\line(1,0){0.15}}
\put(0.85,0){\makebox(0,0){$\times$}} \put(0,0){\circle*{0.15}}}

\newcommand{\PotintS}
{\put(0,0){\dash{4}} \put(1,0){\makebox(0,0){$\times$}}
\put(0,0){\circle*{0.1}}}

\newcommand{\PotintL}
{\put(-1.25,0)\dashH
\put(-1.35,0){\makebox(0,0){x}}
\put(0,0){\circle*{0.15}}}

\newcommand{\potintL}
{\multiput(0.05,0)(0.25,0){3}{\line(-1,0){0.15}}
\put(-0.85,0){\makebox(0,0){x}} \put(0,0){\circle*{0.15}}}

\newcommand{\Effpot}
{\put(0,0)\dashH \put(1.35,0){\makebox(0,0){$\times$}}
\put(1.35,0){\Circle{0.4}}
}

\newcommand{\EffpotQED}
{\linethickness{0.3mm}\put(0,0)\dashH
\put(1.35,0){\makebox(0,0){x}}
\put(1.35,0){\circle{0.4}}\put(1.35,0){\circle{0.45}}\put(1.35,0){\circle{0.5}}
\put(1.35,0){\circle{0.55}}\put(1.35,0){\circle{0.6}}
\put(0,0){\circle*{0.20}} }

\newcommand{\effpot}
{\multiput(0.05,0)(0.25,0){3}{\line(1,0){0.15}}
\put(0.85,0){\makebox(0,0){x}} \put(0.85,0){\Circle{0.3}}
\put(0,0){\Circle*{0.1}}}

\newcommand{\EffpotL}
{\put(-1.25,0)\dashH \put(-1.25,0){\makebox(0,0){x}}
\put(-1.25,0){\circle{0.4}}
}

\newcommand{\TriangUp}[1]
{\put(-0.7,0){\line(1,0){1.4}} \put(0,0.7){\line(-1,-1){0.7}}
\put(0,0.7){\line(1,-1){0.7}} \put(0,0.25){\makebox(0,0){#1}}}

\newcommand{\TriangDn}[1]
{\put(-0.7,0){\line(1,0){1.4}} \put(0,-0.7){\line(1,1){0.7}}
\put(0,-0.7){\line(-1,+1){0.7}} \put(0,-0.25){\makebox(0,0){#1}}}

\newcommand{\Triang}
{\put(0,0){\line(2,1){0.5}}
\put(0,0){\line(2,-1){0.5}}
\put(0.5,-0.25){\line(0,1){0.5}}}

\newcommand{\Triangle}
{\put(0,0){\line(2,1){0.7}} \put(0,0){\line(2,-1){0.7}}
\put(0.7,-0.35){\line(0,1){0.7}}}

\newcommand{\TriangL}
{\put(0,0){\line(-2,1){0.5}}
\put(0,0){\line(-2,-1){0.5}}
\put(-0.5,-0.25){\line(0,1){0.5}}}

\newcommand{\hfint}
{\put(0,0)\dashH
\put(1.25,0){\makebox(0,0){\Triang}}
\put(0,0){\circle*{0.15}}}

\newcommand{\hfintL}
{\put(-1.25,0)\dashH \put(-1.25,0){\makebox(0,0){\TriangL}}
\put(0,0){\circle*{0.15}}}

\newcommand{\Oval}[2]
{\put(0.0,0){\oval(#1,#2)}
\put(0.04,0){\oval(#1,#2)}\put(-0.04,0){\oval(#1,#2)}
\put(0.0,0.04){\oval(#1,#2)}\put(0.0,-0.04){\oval(#1,#2)}
\put(-0.02,0){\oval(#1,#2)}\put(0.02,0){\oval(#1,#2)}
\put(0,0.02){\oval(#1,#2)}\put(0,0,02){\oval(#1,#2)} }

\newcommand{\LoopT}[1]
{\put(0,0){\circle{1}}\put(0,0.02){\circle{1}}\put(0,-0.025){\circle{1}}
\put(-0.02,0){\circle{1}}\put(0.02,0){\circle{1}} 
\put(0.85,0){\makebox(0,0){#1}}}

\newcommand{\LoopTh}[2]
{\put(0,0){\circle{#1}}\put(0,0.02){\circle{#1}}\put(0,-0.025){\circle{#1}}
\put(-0.02,0){\circle{#1}}\put(0.02,0){\circle{#1}} } 

\newcommand{\VPloop}[2]
{\linethickness{0.4mm}\put(0,0){\LoopTh{#1}{#2}}
 \put(0.5,0.05){\VectorDn}
\put(0.85,0){\makebox(0,0){#2}}}

\newcommand{\VPloopt}[1]
{\put(0,0){\circle{1}} \put(0.44,0){\VectorDn}
\put(0.75,0){\makebox(0,0){$#1$}}}

\newcommand{\VPloopL}[1]
{\linethickness{0.4mm}\put(0,0){\LoopT{#1}}
\put(-0.5,-0.05){\VectorUp} \put(-0.75,0){\makebox(0,0){$#1$}}}

\newcommand{\VPloopLt}[1]
{\put(0,0){\circle{1}} \put(-0.46,-0.05){\VectorUp}
\put(-0.75,0){\makebox(0,0){$#1$}}}

\newcommand{\VPloopLR}[2]
{\linethickness{0,3mm}\put(0,0){\LoopT{#1}} \put(-0.5,0){\VectorDn}
\put(0.5,0){\VectorUp} \put(-0.75,0){\makebox(0,0){$#1$}}
\put(0.75,0){\makebox(0,0){$#2$}}}

\newcommand{\VPloopLRt}[2]
{\put(0,0){\circle{1}}
\put(-0.5,0){\VectorDn}
\put(0.5,0){\VectorUp}
\put(-0.75,0){\makebox(0,0){$#1$}}
\put(0.75,0){\makebox(0,0){$#2$}}}

\newcommand{\VPloopD}[2]
{\linethickness{0,3mm}\put(0,0){\LoopT{#1}} \put(0,0.48){\VectorR}
\put(0,-0.48){\Vector} \put(0,0.8){\makebox(0,0){$#1$}}
\put(0,-0.8){\makebox(0,0){$#2$}}}

\newcommand{\VPloopDt}[2]
{\put(0,0){\circle{1}} \put(0,0.4){\VectorR}
\put(0.05,-0.4){\Vector} \put(0,0.8){\makebox(0,0){$#1$}}
\put(0,-0.8){\makebox(0,0){$#2$}}}

\newcommand{\Loop}[2]
{\put(0,0){\oval(0.6,1.25)}\put(0.01,0.01){\oval(0.6,1.25)}\put(-0.01,-0.01){\oval(0.6,1.25)}
\put(0.3,0){\VectorUp}
\put(-0.3,0){\VectorDn}
\put(-0.65,0){\makebox(0,0){$#1$}}
\put(0.65,0){\makebox(0,0){$#2$}}}

\newcommand{\HFexch}[1]
{\put(0,0)\dashH
\qbezier(0,0.01)(0.625,0.515)(1.25,0.015)
\qbezier(0,-0.01)(0.625,0.485)(1.25,-0.015)
\put(0.625,0.26){\Vector}
\put(0.625,0.5){\makebox(0,0){$#1$}}
\put(0,0){\circle*{0.15}}
\put(1.25,0){\circle*{0.15}}}

\newcommand{\HFexcht}[1]
{\put(0,0)\dashH \qbezier(0,0.01)(0.625,0.515)(1.25,0.015)
\put(0.625,0.26){\Vector} \put(0.625,0.5){\makebox(0,0){$#1$}}
\put(0,0){\circle*{0.15}} \put(1.25,0){\circle*{0.15}}}

\newcommand{\dcircH}[2]
{\put(-0.5,0){\multiput(0.05,0)(0.25,0){8}{\line(1,0){0.15}}}
\put(-1,0){\makebox(0,0){$#1$}} \put(2,0){\makebox(0,0){$#2$}}
\put(0,0){\circl} \put(1,0){\circl}}

\newcommand{\dcirc}[2]
{\put(-0.5,0){\multiput(0.05,0)(0.25,0){12}{\line(1,0){0.15}}}
\put(-1,0){\makebox(0,0){$#1$}} \put(3,0){\makebox(0,0){$#2$}}
\put(0,0){\circl} \put(2,0){\circl}}

\newcommand{\dcircl}[2]
{\put(-0.5,0){\makebox(0,0){$#1$}} \put(2.5,0){\makebox(0,0){$#2$}}
\put(0,0){\circl} \put(2,0){\circl}}

\newcommand{\dcircOut}[5]
{\put(0,0){\dcirc{#1}{#2}} \put(0,0){\ellineP{#3}{2}{#4}{}{}{#5}}}

\newcommand{\dcircIn}[5]
{\put(0,0){\dcirc{#1}{#2}}
\put(0,-#3){\ellineP{#3}{2}{#4}{}{}{#5}}}
  
\newcommand{\dcircT}[2]
{\put(-0.5,0){\multiput(0.05,0)(0.25,0){16}{\line(1,0){0.15}}}
\put(-1,0){\makebox(0,0){$#1$}} \put(4,0){\makebox(0,0){$#2$}}
\put(0,0){\circl} \put(3,0){\circl}}

\newcommand{\dcircTIn}[4]
{\put(0,0){\dcircT{#1}{#2}}
\put(0,-1){\Elline{1}{0.5}{}{}}\put(3,-1){\Elline{1}{0.5}{}{}}
\put(-0.4,-0.5){\makebox(0,0){$#3$}}
\put(3.5,-0.5){\makebox(0,0){$#4$}}}

\newcommand{\dcircF}[2]
{\put(-0.5,0){\multiput(0.05,0)(0.25,0){20}{\line(1,0){0.15}}}
\put(-1,0){\makebox(0,0){$#1$}} \put(4,0){\makebox(0,0){$#2$}}
\put(0,0){\circl} \put(4,0){\circl}}

\newcommand{\dcircFIn}[4]
{\put(0,0){\dcircF{$#1$}{$#2$}}
\put(0,-1){\Elline{1}{0.5}{}{}}\put(4,-1){\Elline{1}{0.5}{}{}}
\put(-0.4,-0.5){\makebox(0,0){$#3$}}
\put(4.5,-0.5){\makebox(0,0){$#4$}}}

\newcommand{\dcircFem}[2]
{\put(-0.5,0){\multiput(0.05,0)(0.25,0){24}{\line(1,0){0.15}}}
\put(-1,0){\makebox(0,0){$#1$}} \put(5,0){\makebox(0,0){$#2$}}
\put(0,0){\circl} \put(5,0){\circl}}

\newcommand{\photonPP}
{\qbezier(0,0)(0.08333,0.125)(0.1666667,0)
\qbezier(0.1666667,0)(0.25,-0.125)(0.3333333,0)
\qbezier(0.3333333,0)(0.416667,0.125)(0.5,0)

\qbezier(0.5,0)(0.583333,-0.125)(0.666667,0)
\qbezier(0.666667,0)(0.75,0.125)(0.833333,0)
\qbezier(0.833333,0)(0.916667,-0.125)(1,0)}

\newcommand{\photonnh}
{\qbezier(0,0)(0.08333,0.125)(0.1666667,0)
\qbezier(0.1666667,0)(0.25,-0.125)(0.3333333,0)
\qbezier(0.3333333,0)(0.416667,0.125)(0.5,0)

\qbezier(0.5,0)(0.583333,-0.125)(0.666667,0)
\qbezier(0.666667,0)(0.75,0.125)(0.833333,0)
\qbezier(0.833333,0)(0.916667,-0.125)(1,0) }

\newcommand{\photonnH}
{\qbezier(0,0)(0.08333,0.125)(0.1666667,0)
\qbezier(0.1666667,0)(0.25,-0.125)(0.3333333,0)
\qbezier(0.3333333,0)(0.416667,0.125)(0.5,0)

\qbezier(0.5,0)(0.583333,-0.125)(0.666667,0)
\qbezier(0.666667,0)(0.75,0.125)(0.833333,0)
\qbezier(0.833333,0)(0.916667,-0.125)(1,0)

\qbezier(1,0)(1.083333,0.125)(1.166667,0)
\qbezier(1.166667,0)(1.25,-0.125)(1.333333,0)
\qbezier(1.333333,0)(1.416667,0.125)(1.5,0)}

\newcommand{\photonn}
{\qbezier(0,0)(0.08333,0.125)(0.1666667,0)
\qbezier(0.1666667,0)(0.25,-0.125)(0.3333333,0)
\qbezier(0.3333333,0)(0.416667,0.125)(0.5,0)

\qbezier(0.5,0)(0.583333,-0.125)(0.666667,0)
\qbezier(0.666667,0)(0.75,0.125)(0.833333,0)
\qbezier(0.833333,0)(0.916667,-0.125)(1,0)

\qbezier(1,0)(1.083333,0.125)(1.166667,0)
\qbezier(1.166667,0)(1.25,-0.125)(1.333333,0)
\qbezier(1.333333,0)(1.416667,0.125)(1.5,0)

\qbezier(1.5,0)(1.583333,-0.125)(1.666667,0)
\qbezier(1.666667,0)(1.75,0.125)(1.833333,0)
\qbezier(1.833333,0)(1.916667,-0.125)(2,0) }

\newcommand{\photonH}[3]
{\put(0,0){\photonPP}\put(1,0){\photonP}
\put(0.75,-0.15){\VectorR} \put(0.75,0.35){\makebox(0,0){$#1$}}
\put(0,0){\circl} \put(1.5,0){\circl}
\put(-0.5,0){\makebox(0,0){#2}} \put(2,0){\makebox(0,0){#3}}}

\newcommand{\photon}[3]
{\photonn\put(1,-0.05){\VectorR}
\put(1,0.35){\makebox(0,0.1){$#1$}} \put(0,0){\circl}	
\put(2,0){\circl} \put(-0.5,0){\makebox(0,0){#2}}
\put(2.5,0){\makebox(0,0){#3}}}

\newcommand{\photonp}[3]
{\photonn \put(1,0.45){\makebox(0,0.1){$#1$}}
\put(-0.5,0){\makebox(0,0){#2}} \put(2.5,0){\makebox(0,0){#3}}}

\newcommand{\Photon}[3]
{
\linethickness{0.4mm} \put(0.,-0.00){\photonn{}{}{}}
 \put(0,0){\circl}\put(2,0){\circl}
}

\newcommand{\PPhoton}[3]
{\linethickness{0.7mm} \put(0.,-0.00){\photonn{}{}{}} }

\newcommand{\PPhotonH}[3]
{ \linethickness{0.7mm} \put(0.,-0.00){\photonnH{}{}{}} }

\newcommand{\PhotonH}[3]
{\put(0.,0.06){\photonnH{}{}{}}\put(0.,0.02){\photonnH{}{}{}}
\put(0.,-0.02){\photonnH{}{}{}}\put(0.,-0.06){\photonnH{}{}{}} }

\newcommand{\Photonh}[3]
{\put(0.,0.075){\photonnh{}{}{}}\put(0.,0.025){\photonnh{}{}{}}
\put(0.,-0.025){\photonnh{}{}{}}\put(0.,-0.075){\photonnh{}{}{}} }

\newcommand{\PhotonT}[3]
{\put(0.,0.){\photon{}{}{}} \put(1.,0){\photon{}{}{}}}

\newcommand{\photonT}[3]
{\put(0,0){\photonn{}{#2}{}}\put(1,0){\photonn{}{}{#3}}
\put(1.5,-0.05){\VectorR} \put(1.5,0.35){\makebox(0,0.1){$#1$}}
\put(0,0){\circl} \put(3,0){\circl}
\put(-0.5,0){\makebox(0,0){#2}} \put(3.5,0){\makebox(0,0){#3}}}

\newcommand{\photonF}[3]
{\put(0,0){\photonn{}{}{}}\put(2,0){\photonn{}{}{}}
\put(2,-0.05){\VectorR} \put(1,0.35){\makebox(0,0.1){$#1$}}
\put(0,0){\circl} \put(4,0){\circl}
\put(-0.5,0){\makebox(0,0){#2}} \put(4.5,0){\makebox(0,0){#3}}}

\newcommand{\photonHS}[4]
{\qbezier(0,0)(0.08333,0.125)(0.1666667,0)
\qbezier(0.1666667,0)(0.25,-0.125)(0.3333333,0)
\qbezier(0.3333333,0)(0.416667,0.125)(0.5,0)
\qbezier(0.5,0)(0.583333,-0.125)(0.666667,0)
\qbezier(0.666667,0)(0.75,0.125)(0.833333,0)
\qbezier(0.833333,0)(0.916667,-0.125)(1,0)
\put(0.5,-0.05){\VectorR} \put(0.5,0.35){\makebox(0,0){$#1$}}
\put(0,0){\circl}\put(1,0){\circl} \put(0,-0.5){\makebox(0,0){#2}}
\put(1,-0.5){\makebox(0,0){#3}}}

\newcommand{\photonNEn}[0]
{\qbezier(0,0)(0.22,-0.02)(0.2,0.2)
\qbezier(0.2,0.2)(0.18,0.42)(0.4,0.4)
\qbezier(0.4,0.4)(0.62,0.38)(0.6,0.6)
\qbezier(0.6,0.6)(0.58,0.82)(0.8,0.8)
\qbezier(0.8,0.8)(1.02,0.78)(1,1)
\qbezier(1,1)(0.98,1.22)(1.2,1.2)
\qbezier(1.2,1.2)(1.42,1.18)(1.4,1.4)
\qbezier(1.4,1.4)(1.38,1.62)(1.6,1.6)
\qbezier(1.6,1.6)(1.82,1.58)(1.8,1.8)
\qbezier(1.8,1.8)(1.78,2.02)(2,2) }

\newcommand{\photonNEu}[0]
{\qbezier(0,0)(-0.02,0.22)(0.2,0.2)
\qbezier(0.2,0.2)(0.42,0.18)(0.4,0.4)
\qbezier(0.4,0.4)(0.38,0.62)(0.6,0.6)
\qbezier(0.6,0.6)(0.82,0.58)(0.8,0.8)
\qbezier(0.8,0.8)(0.78,1.02)(1,1)
\qbezier(1,1)(1.22,0.98)(1.2,1.2)
\qbezier(1.2,1.2)(1.18,1.42)(1.4,1.4)
\qbezier(1.4,1.4)(1.62,1.38)(1.6,1.6)
\qbezier(1.6,1.6)(1.58,1.82)(1.8,1.8)
\qbezier(1.8,1.8)(2.02,1.78)(2,2) }

\newcommand{\photonNE}[3]
{\photonNEn \put(1,1){\makebox(0.05,-0.2){\VectorUp}}
\put(0,0){\circl} \put(2,2){\circl}
\put(0.8,0.6){\makebox(0,0){$#1$}}
\put(-0.35,-1){\makebox(0,2){$#2$}}
\put(2.35,1){\makebox(0,2){$#3$}}}

\newcommand{\photonNNE}[3]
{\qbezier(0,0)   (0.28,-0.02)(0.2,0.3)
\qbezier(0.2,0.3)(0.12,0.52)(0.4,0.6)
\qbezier(0.4,0.6)(0.68,0.58)(0.6,0.9)
\qbezier(0.6,0.9)(0.52,1.12)(0.8,1.2)
\qbezier(0.8,1.2)(1.08,1.18)(1,1.5) \qbezier(1,1.5)
(0.92,1.72)(1.2,1.8) \qbezier(1.2,1.8)(1.48,1.86)(1.4,2.1)
\qbezier(1.4,2.1)(1.365,2.24)(1.6,2.4)
\qbezier(1.6,2.4)(1.835,2.46)(1.8,2.7)
\qbezier(1.8,2.7)(1.765,2.84)(2,3)
\put(0.6,0.8){\makebox(0,0){\VectorUp}} \put(0,0){\circle*{0.15}}
\put(2,3){\circle*{0.15}} \put(1,0.8){\makebox(0,0){$#1$}}
\put(-0.35,0){\makebox(0,0){$#2$}}
\put(2.35,3){\makebox(0,0){$#3$}}}

\newcommand{\photonENE}[3]
{\qbezier(0,0)(0.17,-0.04)(0.2,0.1)
\qbezier(0.2,0.1)(0.23,0.32)(0.4,0.2)
\qbezier(0.4,0.2)(0.57,0.16)(0.6,0.3)
\qbezier(0.6,0.3)(0.63,0.52)(0.8,0.4)
\qbezier(0.8,0.4)(0.97,0.36)(1,0.5)
\qbezier(1,0.5)(1.03,0.72)(1.2,0.6)
\qbezier(1.2,0.6)(1.37,0.56)(1.4,0.7)
\qbezier(1.4,0.7)(1.43,0.92)(1.6,0.8)
\qbezier(1.6,0.8)(1.77,0.76)(1.8,0.9)
\qbezier(1.8,0.9)(1.83,1.12)(2,1)
\put(1,0.8){\makebox(0,0){\VectorR}} \put(0,0){\circle*{0.15}}
\put(2,1){\circle*{0.15}} \put(1,1){\makebox(0,-0){$#1$}}
\put(-0.35,-1){\makebox(0,2){$#2$}}
\put(2.35,0){\makebox(0,2){$#3$}}}

\newcommand{\photonENEh}[3]
{\qbezier(0,0)(0.17,-0.04)(0.2,0.1)
\qbezier(0.2,0.1)(0.23,0.32)(0.4,0.2)
\qbezier(0.4,0.2)(0.57,0.16)(0.6,0.3)
\qbezier(0.6,0.3)(0.63,0.52)(0.8,0.4)
\qbezier(0.8,0.4)(0.97,0.36)(1,0.5)
\put(1,0.85){\makebox(0,-0){$#1$}}
\put(-0.35,-1){\makebox(0,2){$#2$}}
\put(2.35,0){\makebox(0,2){$#3$}}}

\newcommand{\photonNWn}[0]
{\qbezier(0,0)(-0.22,-0.02)(-0.2,0.2)
\qbezier(-0.2,0.2)(-0.18,0.42)(-0.4,0.4)
\qbezier(-0.4,0.4)(-0.62,0.38)(-0.6,0.6)
\qbezier(-0.6,0.6)(-0.58,0.82)(-0.8,0.8)
\qbezier(-0.8,0.8)(-1.02,0.78)(-1,1) \qbezier(-1,1)
(-0.98,1.22)(-1.2,1.2) \qbezier(-1.2,1.2)(-1.42,1.18)(-1.4,1.4)
\qbezier(-1.4,1.4)(-1.38,1.62)(-1.6,1.6)
\qbezier(-1.6,1.6)(-1.82,1.58)(-1.8,1.8)
\qbezier(-1.8,1.8)(-1.78,2.02)(-2,2)}

\newcommand{\photonNWu}[0]
{\qbezier(0,0)(0.02,-0.22)(-0.2,0.2)
\qbezier(-0.2,0.2)(-0.42,0.18)(-0.4,0.4)
\qbezier(-0.4,0.4)(-0.38,0.62)(-0.6,0.6)
\qbezier(-0.6,0.6)(-0.82,0.58)(-0.8,0.8)
\qbezier(-0.8,0.8)(-0.78,1.02)(-1,1)
\qbezier(-1,1)(-1.22,0.98)(-1.2,1.2)
\qbezier(-1.2,1.2)(-1.18,1.42)(-1.4,1.4)
\qbezier(-1.4,1.4)(-1.62,1.38)(-1.6,1.6)
\qbezier(-1.6,1.6)(-1.58,1.8)(-1.8,1.8)
\qbezier(-1.8,1.8)(-2.02,1.78)(-2,2)}

\newcommand{\photonNW}[3]
{\put(0,0)\photonNWn \put(-1,1){\makebox(0,-0.2){\VectorUp}}
\put(0,0){\circl} \put(-2,2){\circl}
\put(-1.1,1.4){\makebox(0,0){$#1$}}
\put(-2.35,2){\makebox(0,0){$#2$}}
\put(0.35,0){\makebox(0,0){$#3$}}}

\newcommand{\photonTNE}[3]
{\photonNEn \put(1,1)\photonNEu
\put(1.5,1.5){\makebox(0.05,-0.2){\VectorUp}} \put(0,0){\circl}
\put(3,3){\circl} \put(1,1){\makebox(-0.4,1){$#1$}}
\put(-0.35,0){\makebox(0,0){$#2$}}
\put(3.35,3.2){\makebox(0,0){$#3$}}}

\newcommand{\photonTNW}[3]
{\photonNWn \put(-1,1)\photonNWu
\put(-1.5,1.5){\makebox(0.05,-0.2){\VectorUp}} \put(0,0){\circl}
\put(-3,3){\circl} \put(1,1){\makebox(-0.4,1){$#1$}}
\put(-0.35,-1){\makebox(0,2){$#2$}}
\put(3.35,1){\makebox(0,2){$#3$}}}

\newcommand{\photonFNW}[3]
{\photonNWn \put(-2,2)\photonNWu
\put(-2,2){\makebox(0.05,-0.2){\VectorUp}} \put(0,0){\circl}
\put(-4,4){\circl} \put(1,1){\makebox(-0.4,1){$#1$}}
\put(-0.35,-1){\makebox(0,2){$#2$}}
\put(4.35,1){\makebox(0,2){$#3$}}}

\newcommand{\photonSEst}[3]
{\qbezier(0,0)(-0.22,-0.02)(-0.2,0.2)
\qbezier(-0.2,0.2)(-0.18,0.42)(-0.4,0.4)
\qbezier(-0.4,0.4)(-0.62,0.38)(-0.6,0.6)
\qbezier(-0.6,0.6)(-0.58,0.82)(-0.8,0.8)
\qbezier(-0.8,0.8)(-1.02,0.78)(-1,1) \qbezier(-1,1)
(-0.98,1.22)(-1.2,1.2) \qbezier(-1.2,1.2)(-1.42,1.18)(-1.4,1.4)
\qbezier(-1.4,1.4)(-1.38,1.62)(-1.6,1.6)
\qbezier(-1.6,1.6)(-1.82,1.58)(-1.8,1.8)
\qbezier(-1.8,1.8)(-1.78,2.02)(-2,2)
\put(-1,1){\makebox(0,0){\VectorDn}} \put(0,0){\circle*{0.15}}
\put(-2,2){\circle*{0.15}} \put(-1,1){\makebox(0.4,0.7){$#1$}}
\put(-2.35,2){\makebox(0,0){$#2$}}
\put(0.35,0){\makebox(0,0){$#3$}}}

\newcommand{\photonWNW}[3]
{\qbezier(0,0)(-0.17,-0.04)(-0.2,0.1)
\qbezier(-0.2,0.1)(-0.23,0.32)(-0.4,0.2)
\qbezier(-0.4,0.2)(-0.57,0.16)(-0.6,0.3)
\qbezier(-0.6,0.3)(-0.63,0.52)(-0.8,0.4)
\qbezier(-0.8,0.4)(-0.97,0.36)(-1,0.5)
\qbezier(-1,0.5)(-1.03,0.72)(-1.2,0.6)
\qbezier(-1.2,0.6)(-1.37,0.56)(-1.4,0.7)
\qbezier(-1.4,0.7)(-1.43,0.92)(-1.6,0.8)
\qbezier(-1.6,0.8)(-1.77,0.76)(-1.8,0.9)
\qbezier(-1.8,0.9)(-1.83,1.12)(-2,1)
\put(-1.1,0.55){\makebox(0,0){$\;$\Vector}}
\put(0,0){\circle*{0.15}} \put(-2,1){\circle*{0.15}}
\put(-1,0.5){\makebox(0.6,0.65){$#1$}}
\put(0.35,-1){\makebox(0,2){$#3$}}
\put(-2.35,0){\makebox(0,2){$#2$}}}

\newcommand{\Crossphotons}[6]
{\put(0,0){\photonNE{#5}{}{}}
\put(2,0){\photonNW{#6}{}{}}
\put(-0.35,0){\makebox(0,0){$#1$}}
\put(2.35,2){\makebox(0,0){$#2$}}
\put(2.35,0){\makebox(0,0){$#3$}}
\put(-0.35,2){\makebox(0,0){$#4$}}
}

\newcommand{\photonNe}[3]
{\qbezier(0,0)(0.22,-0.02)(0.2,0.2)
\qbezier(0.2,0.2)(0.18,0.42)(0.4,0.4)
\qbezier(0.4,0.4)(0.62,0.38)(0.6,0.6)
\qbezier(0.6,0.6)(0.58,0.82)(0.8,0.8)
\qbezier(0.8,0.8)(1.02,0.78)(1,1)
\qbezier(1,1)(0.98,1.22)(1.2,1.2)
\put(0.75,0.75){\makebox(-0.6,0.4){$#1$}}
\put(-0.35,0){\makebox(0,0){$#2$}}
\put(1.85,1.5){\makebox(0,0){$#3$}}}

\newcommand{\PhotonNe}[3]
{\put(0,0){\photonNe{}{}{}}\put(0.05,-0.05){\photonNe{}{}{}}
\put(-0.05,0.05){\photonNe{}{}{}}
\put(0.75,0.75){\makebox(-0.6,0.4){$#1$}}
\put(-0.35,0){\makebox(0,0){$#2$}}
\put(1.85,1.5){\makebox(0,0){$#3$}}}

\newcommand{\photonNw}[3]
{\qbezier(0,0)(0.02,0.22)(-0.2,0.2)
\qbezier(-0.2,0.2)(-0.42,0.18)(-0.4,0.4)
\qbezier(-0.4,0.4)(-0.38,0.62)(-0.6,0.6)
\qbezier(-0.6,0.6)(-0.82,0.58)(-0.8,0.8)
\qbezier(-0.8,0.8)(-0.78,1.02)(-1,1) \qbezier(-1,1)
(-1.22,0.98)(-1.2,1.2) \qbezier(-1.2,1.2)(-1.18,1.42)(-1.4,1.4)
\qbezier(-1.4,1.4)(-1.5,1.38)(-1.5,1.5)
\put(0,0){\circle*{0.20}}
\put(-0.7,0.7){\makebox(1,0.3){$#1$}}
\put(0.35,0){\makebox(0,0){$#3$}}
\put(-1.85,1.5){\makebox(0,0){$#2$}} }

\newcommand{\elstat}[3]
{\multiput(0.06,0)(0.35,0){6}{\line(1,0){0.15}}
\put(1,0.35){\makebox(0,0){$#1$}} \put(-0.35,0){\makebox(0,0){#2}}
\put(0,0){\circl}\put(2,0){\circl}
 \put(2.35,0){\makebox(0,0){#3}}}

\newcommand{\PairOut}[6]
 {\Pair{#1}{#2}{#3}\put(0,0){\elline{#4}{#5}{}}
 \put(#1,0){\elline{#4}{}{#6}}}

\newcommand{\Pair}[3]
{{\linethickness{1mm} \put(0,0){\line(1,0){#1}}}
 \put(-0.3,0){\makebox(0,0){$#2$}}
\put(0,0){\circl}\put(#1,0){\circl}
 \put(#1,0){\makebox(0.6,0){$#3$}}}

\newcommand{\PPair}[3]
{\linethickness{2mm} \put(0,0){\line(1,0){#1}}
 \put(-0.3,0){\makebox(0,0){$#2$}}
 \put(#1,0){\makebox(0.6,0){$#3$}}}

\newcommand{\OmQED}[3]
{\linethickness{2mm} \put(0,0){\line(1,0){#1}}
 \put(-0.3,0){\makebox(0,0){$#2$}}
 \put(#1,0){\makebox(0.6,0){$#3$}}}

\newcommand{\OmQEDP}[3]
{\linethickness{2mm} \put(0,0){\line(1,0){#1}}
 \put(-0.3,0){\makebox(0,0){$#2$}}
 \put(#1,0){\makebox(0.6,0){$#3$}}}

\newcommand{\PPairP}
{\linethickness{2mm} \put(0,0){\line(1,0){2}}
}

\newcommand{\PPairH}
{\linethickness{1.5mm} \put(0,0){\line(1,0){1.5}}
}

\newcommand{\PPairh}
{\linethickness{1.5mm} \put(0,0){\line(1,0){1}}
}

\newcommand{\UQED}
{\put(0,-0.125){\Elstat{}{}{}}\put(0,0.15){\Elstat{}{}{}}
\put(0.,0.){\Photon{}{}{}}}

\newcommand{\UQEDH}
{\put(0,-0.125){\ElstatH{}{}{}}\put(0,0.15){\ElstatH{}{}{}}
\put(0.,0.){\PhotonH{}{}{}}}

\newcommand{\UQEDh}
{\put(0,-0.125){\Elstath{}{}{}}\put(0,0.15){\Elstath{}{}{}}
\put(0.,0.){\Photonh{}{}{}}}

\newcommand{\Usp}
{{\linethickness{0.3mm}\put(0,-0.125){\line(1,0){2}}\put(0,0.125){\line(1,0){2}}}
\put(0.,0.){\photonn{}{}{}}}

\newcommand{\UspH}
{{\linethickness{0.3mm}\put(0,-0.125){\line(1,0){1.5}}\put(0,0.125){\line(1,0){1.5}}}
\put(0.,0.){\photonnH{}{}{}}}

\newcommand{\Usph}
{{\linethickness{0.3mm}\put(0,-0.125){\line(1,0){1}}\put(0,0.125){\line(1,0){1}}}
\put(0.,0.){\photonnh{}{}{}}}

\newcommand{\PairQEDT}
{\put(0,0){\PairQED}\put(1.,0){\PairQED}}

\newcommand{\PairQEDH}
{\linethickness{0.2mm}\put(0,-0.125){\line(1,0){1.5}}\put(0,0.125){\line(1,0){1.5}}
\put(0.,0.){\PhotonH{}{}{}}}

\newcommand{\PairQEDh}
{\put(0,0){\Pair{1}{}{}}\put(0.,0.1){\photonnh{}{}{}}\put(0.,-0.1){\photonnh{}{}{}}
}

\newcommand{\PairIn}[6]
 {\Pair{#1}{#2}{#3}\put(0,-#4){\elline{#4}{#5}{}}
 \put(#1,-#4){\elline{#4}{}{#6}}}

\newcommand{\CoulIn}[5]
 {\Elstat{}{#1}{#2}\put(0,-#3){\elline{#3}{#4}{}}
 \put(2,-#3){\elline{#3}{}{#5}}}

\newcommand{\CoulInT}[5]
 {\ElstatT{}{#1}{#2}\put(0,-#3){\elline{#3}{#4}{}}
 \put(3,-#3){\elline{#3}{}{#5}}}

\newcommand{\CoulInF}[5]
 {\ElstatF{}{#1}{#2}\put(0,-#3){\elline{#3}{#4}{}}
 \put(4,-#3){\elline{#3}{}{#5}}}

 \newcommand{\CoulInh}[5]
 {\Elstath{}{#1}{#2}\put(0,-#3){\elline{#3}{#4}{}}
 \put(1,-#3){\elline{#3}{}{#5}}}

\newcommand{\elstatn}[3]
{\multiput(0.06,0)(0.25,0){8}{\line(1,0){0.15}}
\put(1,0.35){\makebox(0,0){$#1$}} \put(-0.35,0){\makebox(0,0){#2}}
 \put(2.35,0){\makebox(0,0){#3}}}

\newcommand{\Elstat}[3]
{\linethickness{0.3mm}\multiput(0.06,0)(0.35,0){6}{\line(1,0){0.15}}
\put(1,0.35){\makebox(0,0){$#1$}}
 \put(0,0){\circl}\put(2,0){\circl}
 \put(-0.35,0){\makebox(0,0){#2}}
\put(2.35,0){\makebox(0,0){#3}}}

\newcommand{\Elstatn}[3]
{\linethickness{0.3mm}\multiput(0.06,0)(0.35,0){6}{\line(1,0){0.15}}
\put(1,0.35){\makebox(0,0){$#1$}}}

\newcommand{\elstatT}[3]
{\linethickness{0.4mm}\multiput(0.06,0)(0.333,0){9}{\line(1,0){0.15}}
\put(1,0.35){\makebox(0,0){$#1$}} \put(0,0){\circl}
\put(3,0){\circl} \put(-0.35,0){\makebox(0,0){#2}}
\put(3.35,0){\makebox(0,0){#3}}}

\newcommand{\ElstatT}[3]
{\linethickness{0.4mm}\multiput(0.06,0)(0.333,0){9}{\line(1,0){0.15}}
\put(1,0.35){\makebox(0,0){$#1$}} \put(0,0){\circl}
\put(3,0){\circl} \put(-0.35,0){\makebox(0,0){#2}}
\put(3.35,0){\makebox(0,0){#3}}}

\newcommand{\ElstatF}[3]
{\linethickness{0.4mm}\multiput(0.06,0)(0.333,0){12}{\line(1,0){0.15}}
\put(1,0.35){\makebox(0,0){$#1$}} \put(0,0){\circl}
\put(4,0){\circl} \put(-0.35,0){\makebox(0,0){#2}}
\put(4.35,0){\makebox(0,0){#3}}}

\newcommand{\elstatF}[3]
{\multiput(0.06,0)(0.333,0){12}{\line(1,0){0.15}}
\put(1,0.35){\makebox(0,0){$#1$}} \put(0,0){\circl}
\put(4,0){\circl} \put(-0.35,0){\makebox(0,0){#2}}
\put(4.35,0){\makebox(0,0){#3}}}

\newcommand{\Breit}[3]
{\multiput(0.0,0)(0.4,0){6}{\circle*{0.2}}
\put(1,0.35){\makebox(0,0){$#1$}}
  \put(-0.35,0){\makebox(0,0){#2}}
\put(2.35,0){\makebox(0,0){#3}}}

\newcommand{\Melstat}[1]
{\multiput(0,0)(0,0.2){#1}{\Elstat{}{}{}}}

\newcommand{\MelstatT}[1]
{\multiput(0,0)(0,0.2){#1}{\elstatT{}{}{}}}

\newcommand{\Multiline}[3]
{\linethickness{0.2mm} \put(0,-0.1){\line(1,0){#1}}
 \put(0,0){\line(1,0){#1}}\put(0,0.1){\line(1,0){#1}}
 \put(-0.35,0){\makebox(0,0){#2}}
\put(2.35,0){\makebox(0,0){#3}}}

\newcommand{\elstatH}[3]
{\multiput(0.06,0)(0.25,0){6}{\line(1,0){0.15}}
\put(0.75,0.25){\makebox(0,0){$#1$}} \put(0,0){\circle*{0.1}}
\put(1.5,0){\circle*{0.1}}
\put(-0.35,0){\makebox(0,0){#2}}
\put(2.35,0){\makebox(0,0){#3}}}

\newcommand{\ElstatH}[3]
{\linethickness{0.4mm}\multiput(0.06,0)(0.25,0){6}{\line(1,0){0.15}}
\put(0.75,0.25){\makebox(0,0){$#1$}} \put(0,0){\circl}
\put(1.5,0){\circl}
\put(-0.35,0){\makebox(0,0){#2}} \put(2.35,0){\makebox(0,0){#3}}}

\newcommand{\elstath}[3]
{\multiput(0.06,0)(0.25,0){4}{\line(1,0){0.15}}
\put(0.75,0.25){\makebox(0,0){$#1$}} 
\put(-0.35,0){\makebox(0,0){#2}} \put(2.35,0){\makebox(0,0){#3}}}

\newcommand{\Elstath}[3]
{\linethickness{0.4mm}
\multiput(0.06,0)(0.25,0){4}{\line(1,0){0.15}}
\put(0.75,0.25){\makebox(0,0){$#1$}} \put(0,0){\circle*{0.1}}
\put(1.5,0){\circle*{0.1}}
\put(-0.35,0){\makebox(0,0){#2}} \put(2.35,0){\makebox(0,0){#3}}}

\newcommand{\BreitH}[3]
{\multiput(0.15,0)(0.3,0){5}{\circle*{0.1}}
\put(0.75,0.25){\makebox(0,0){$#1$}} 
\put(-0.35,0){\makebox(0,0){#2}}
\put(2.35,0){\makebox(0,0){#3}}}

\newcommand{\RetBreitDH}[3]
{\small\multiput(0.15,0.3)(0.3,-0.15){5}{\circle*{0.1}}
\put(0.75,0.25){\makebox(0,0){$#1$}}
\put(-0.35,0){\makebox(0,0){#2}} \put(2.35,0){\makebox(0,0){#3}}}

\newcommand{\RetBreitH}[3]
{\small\multiput(0.15,-0.3)(0.3,0.15){5}{\circle*{0.1}}
\put(0.75,0.25){\makebox(0,0){$#1$}}
\put(-0.35,0){\makebox(0,0){#2}} \put(2.35,0){\makebox(0,0){#3}}}

\newcommand{\elsta}[3]
{\multiput(0.06,0)(0.25,0){4}{\line(1,0){0.15}}
\put(1,0.35){\makebox(0,0){$#1$}}
\put(0,0){\circle*{0.1}}
\put(1,0){\circle*{0.1}}
\put(-0.35,0){\makebox(0,0){#2}}
\put(1.35,0){\makebox(0,0){#3}}}

\newcommand{\elstatNO}[3]
{\multiput(0.06,0.08)(0.01,0.01){14}{\tiny.}
\multiput(0.30,0.32)(0.01,0.01){14}{\tiny.}
\multiput(0.55,0.57)(0.01,0.01){14}{\tiny.}
\multiput(0.79,0.81)(0.01,0.01){14}{\tiny.}
\multiput(1.03,1.05)(0.01,0.01){14}{\tiny.}
\multiput(1.27,1.29)(0.01,0.01){14}{\tiny.}
\multiput(1.51,1.53)(0.01,0.01){14}{\tiny.}
\multiput(1.75,1.77)(0.01,0.01){14}{\tiny.}
\put(0.95,0.75){\makebox(0,0){\VectorUr}}
\put(0,0){\circle*{0.15}}
\put(2,2){\circle*{0.15}}
\put(0.75,1.25){\makebox(0,0){$#1$}}
\put(-0.5,0){\makebox(0,0){$#2$}}
\put(2.5,2){\makebox(0,0){$#3$}}
}

\newcommand{\elstatNW}[3]
{\multiput(-0.05,0.05)(-0.015,0.015){10}{\circle*{0.02}}
\multiput(-0.3,0.3)(-0.015,0.015){10}{\circle*{0.02}}
\multiput(-0.55,0.55)(-0.015,0.015){10}{\circle*{0.02}}
\multiput(-0.8,0.8)(-0.015,0.015){10}{\circle*{0.03}}
\multiput(-1.05,1.05)(-0.015,0.015){10}{\circle*{0.03}}
\multiput(-1.3,1.3)(-0.015,0.015){10}{\circle*{0.03}}
\multiput(-1.55,1.55)(-0.015,0.015){10}{\circle*{0.03}}
\multiput(-1.8,1.8)(-0.015,0.015){10}{\circle*{0.03}}
\put(-0.9,0.9){\makebox(0,0){\VectorUl}}
\put(0,0){\circle*{0.215}}
\put(-2,2){\circle*{0.2153}}
\put(-0.9,0.9){\makebox(0,0){\VectorUl}}
\put(0,0){\circle*{0.215}}
\put(-2,2){\circle*{0.215}}
\put(-0.5,1){\makebox(0,0){$#1$}}
\put(0,-0.5){\makebox(0,0){$#2$}}
\put(-2,2.5){\makebox(0,0){$#3$}}
}

\newcommand{\photonSE}[6]
{\put(0,0){\photonHS{#3}{#4}{}{}} \put(1.5,0){\VPloopD{#1}{#2}}
\put(2,0){\photonHS{#5}{#6}{}{}}}

\newcommand{\photonSEt}[5]
{\put(0,0){\photonHS{#3}{#4}{}{}}
\put(1.5,0){\VPloopDt{#1}{#2}}
\put(2,0){\photonHS{#5}{}{}{}}}

\newcommand{\ElSE}[3]
{\qbezier(0,-1)(.2025,-1.1489)(0.3420,-0.9397)
\qbezier(0.3420,-0.9397)(0.4167,-0.7217)(0.6428,-0.766)
\qbezier(0.6428,-0.766)(0.8937,-0.7499)(0.866,-0.5)
\qbezier(0.866,-0.5)(0.7831,-0.2850)(0.9848,-0.1736)
\qbezier(0.9848,-0.1736)(1.1667,0)(0.9848,0.1736)
\qbezier(0.9848,0.1736)(0.7831,0.2850)(0.866,0.5)
\qbezier(0.866,0.5)(0.8937,0.7499)(0.6428,0.766)
\qbezier(0.6428,0.766)(0.4167,0.7217)(0.3420,0.9397)
\qbezier(0.3420,0.9397)(.2025,1.1489)(0,1) \put(1,0.02){\VectorUp}
\put(0,1){\circl} \put(0,-1){\circl}
\put(1.45,0){\makebox(0,0){$#1$}} \put(-0.35,-1){\makebox(0,0){#2}}
\put(-0.35,1){\makebox(0,0){#3}}}

\newcommand{\ElSEN}[3]
{\qbezier(0,-1)(.1718,-1.0420)(0.1863,-0.8399)
\qbezier(0.1863,-0.8399)(0.1704,-0.6495)(0.3357,-0.6346)
\qbezier(0.3357,-0.6346)(0.5055,-0.5831)(0.4399,-0.3950)
\qbezier(0.4399,-0.3950)(0.3517,-0.2330)(0.4935,-0.1341)
\qbezier(0.4935,-0.1341)(0.6252,0)(0.4935,0.1341)
\qbezier(0.4935,0.1341)(0.3517,0.2330)(0.4399,0.3950)
\qbezier(0.4399,0.3950)(0.5055,0.5831)(0.3357,0.6346)
\qbezier(0.3357,0.6346)(0.1704,0.6495)(0.1863,0.8399)
\qbezier(0.1863,0.8399)(0.1718,1.0420)(0,1)
\put(0.5,0.02){\VectorUp} \put(0,1){\circle*{0.15}}
\put(0,-1){\circle*{0.15}} \put(0.85,0){\makebox(0,0){$#1$}}
\put(-0.35,-1){\makebox(0,0){#2}}
\put(-0.35,1){\makebox(0,0){#3}}}

\newcommand{\ElSENR}[3]
{\qbezier(-0.4472,-0.8944)(-0.3123,-1.0088)(-0.209,-0.8346)
\qbezier(-0.2090,-0.8346)(-0.138,-0.6572)(0.0164,-0.7177)
\qbezier(0.01664,-0.7177)(0.1914,-0.7476)(0.2169,-0.55)
\qbezier(0.2169,-0.55)(0.2103,-0.3657)(0.3815,-0.3406)
\qbezier(0.3815,-0.3406)(0.5593,-0.2796)(0.5014,-0.1079)
\qbezier(0.5014,-0.1008)(0.4188,0.0512)(0.5701,0.1565)
\qbezier(0.5701,0.1565)(0.7129,0.2954)(0.584,0.4175)
\qbezier(0.584,0.4175)(0.4429,0.5047)(0.5423,0.6679)
\qbezier(0.5423,0.6679)(0.6617,0.8552)(0.4472,0.8944)
\put(0.5,-0.1){\VectorUp} \put(0.475,0.95){\circle*{0.15}}
\put(-0.475,-0.95){\circle*{0.15}}
\put(0.9,0){\makebox(0,0){$#1$}} \put(-0.85,-1){\makebox(0,0){#2}}
\put(0.15,1){\makebox(0,0){#3}}}

\newcommand{\ElSENL}[3]
{\qbezier(-0.4472,0.8944)(-0.3123,1.0088)(-0.209,0.8346)
\qbezier(-0.2090,0.8346)(-0.138,0.6572)(0.0164,0.7177)
\qbezier(0.01664,0.7177)(0.1914,0.7476)(0.2169,0.55)
\qbezier(0.2169,0.55)(0.2103,0.3657)(0.3815,0.3406)
\qbezier(0.3815,0.3406)(0.5593,0.2796)(0.5014,0.1079)
\qbezier(0.5014,0.1008)(0.4188,-0.0512)(0.5701,-0.1565)
\qbezier(0.5701,-0.1565)(0.7129,-0.2954)(0.584,-0.4175)
\qbezier(0.584,-0.4175)(0.4429,-0.5047)(0.5423,-0.6679)
\qbezier(0.5423,-0.6679)(0.6617,-0.8552)(0.4472,-0.8944)
\put(0.5,0.1){\VectorUp} \put(-0.475,0.95){\circle*{0.15}}
\put(0.475,-0.95){\circle*{0.15}}
\put(-0.9,0){\makebox(0,0){$#1$}} \put(0.85,-1){\makebox(0,0){#2}}
\put(-0.85,1){\makebox(0,0){#3}}}

\newcommand{\ElSEL}[3]
{\qbezier(0,-1)(-.2025,-1.1489)(-0.3420,-0.9397)
\qbezier(-0.3420,-0.9397)(-0.4167,-0.7217)(-0.6428,-0.766)
\qbezier(-0.6428,-0.766)(-0.8937,-0.7499)(-0.866,-0.5)
\qbezier(-0.866,-0.5)(-0.7831,-0.2850)(-0.9848,-0.1736)
\qbezier(-0.9848,-0.1736)(-1.1667,0)(-0.9848,0.1736)
\qbezier(-0.9848,0.1736)(-0.7831,0.2850)(-0.866,0.5)
\qbezier(-0.866,0.5)(-0.8937,0.7499)(-0.6428,0.766)
\qbezier(-0.6428,0.766)(-0.4167,0.7217)(-0.3420,0.9397)
\qbezier(-0.3420,0.9397)(-.2025,1.1489)(0,1)
\put(-1,0.02){\VectorUp} 
\put(-1.45,0){\makebox(0,0){$#1$}}
\put(0.35,-1){\makebox(0,0){#2}} \put(0.35,1){\makebox(0,0){#3}}}

\newcommand{\SEpolt}[5]
{\qbezier(0,-1.5)(.2025,-1.6489)(0.3420,-1.4397)
\qbezier(0.3420,-1.4397)(0.4167,-1.2217)(0.6428,-1.266)
\qbezier(0.6428,-1.266)(0.8937,-1.2499)(0.866,-1)
\qbezier(0.866,-1)(0.7831,-0.7850)(0.9848,-0.6736)
\qbezier(1,-0.5)(1.1,-0.5)(0.9848,-0.6736)
\qbezier(1,0.5)(1.1,0.5)(0.9848,0.6736)
\qbezier(0.9848,0.6736)(0.7831,0.7850)(0.866,1)
\qbezier(0.866,1)(0.8937,1.2499)(0.6428,1.266)
\qbezier(0.6428,1.266)(0.4167,1.2217)(0.3420,1.4397)
\qbezier(0.3420,1.4397)(.2025,1.6489)(0,1.5)
\put(1,0){\VPloopLR{#1}{#2}} \put(0.87,-1){\VectorUp}
\put(0.67,1.23){\Vector} \put(1.3,-1){\makebox(0,0){#3}}
\put(1,0.5){\circle*{0.15}} \put(1,-0.5){\circle*{0.15}}
\put(0,1.5){\circle*{0.15}} \put(0,-1.5){\circle*{0.15}}
\put(-0.35,-1.5){\makebox(0,0){#4}}
\put(-0.35,1.5){\makebox(0,0){#5}}}

\newcommand{\SEpoltNA}[5]
{\qbezier(0,-1.5)(.2025,-1.6489)(0.3420,-1.4397)
\qbezier(0.3420,-1.4397)(0.4167,-1.2217)(0.6428,-1.266)
\qbezier(0.6428,-1.266)(0.8937,-1.2499)(0.866,-1)
\qbezier(0.866,-1)(0.7831,-0.7850)(0.9848,-0.6736)
\qbezier(1,-0.5)(1.1,-0.5)(0.9848,-0.6736)
\qbezier(1,0.5)(1.1,0.5)(0.9848,0.6736)
\qbezier(0.9848,0.6736)(0.7831,0.7850)(0.866,1)
\qbezier(0.866,1)(0.8937,1.2499)(0.6428,1.266)
\qbezier(0.6428,1.266)(0.4167,1.2217)(0.3420,1.4397)
\qbezier(0.3420,1.4397)(.2025,1.6489)(0,1.5)
\put(1,0){\circle{1}}
\put(0.87,-1){\VectorUp}
\put(0.67,1.23){\Vector}
\put(1.3,-1){\makebox(0,0){#3}}
\put(1,0.5){\circle*{0.15}}
\put(1,-0.5){\circle*{0.15}}
\put(0,1.5){\circle*{0.15}}
\put(0,-1.5){\circle*{0.15}}
\put(-0.35,-1.5){\makebox(0,0){#4}}
\put(-0.35,1.5){\makebox(0,0){#5}}}

\newcommand{\SEpol}[5]
{\qbezier(0,-1.5)(.2025,-1.6489)(0.3420,-1.4397)
\qbezier(0.3420,-1.4397)(0.4167,-1.2217)(0.6428,-1.266)
\qbezier(0.6428,-1.266)(0.8937,-1.2499)(0.866,-1)
\qbezier(0.866,-1)(0.7831,-0.7850)(0.9848,-0.6736)
\qbezier(1,-0.5)(1.1,-0.5)(0.9848,-0.6736)
\qbezier(1,0.5)(1.1,0.5)(0.9848,0.6736)
\qbezier(0.9848,0.6736)(0.7831,0.7850)(0.866,1)
\qbezier(0.866,1)(0.8937,1.2499)(0.6428,1.266)
\qbezier(0.6428,1.266)(0.4167,1.2217)(0.3420,1.4397)
\qbezier(0.3420,1.4397)(.2025,1.6489)(0,1.5)
\put(1,0){\VPloopLR{#1}{#2}}
\put(0.87,-1){\VectorUp}
\put(0.67,1.23){\Vector}
\put(1.3,-1){\makebox(0,0){#3}}
\put(1,0.5){\circle*{0.15}}
\put(1,-0.5){\circle*{0.15}}
\put(0,1.5){\circle*{0.15}}
\put(0,-1.5){\circle*{0.15}}
\put(-0.35,-1.5){\makebox(0,0){#4}}
\put(-0.35,1.5){\makebox(0,0){#5}}}

%% file: FigGreen.tex

\newcommand{\SEboxT}[2]
{\put(0,0){\LineH{#1}} \put(0,#2){\LineH{#1}}
\put(0,0){\LineV{#2}} \put(#1,0){\LineV{#2}}
\put(0,0){\line(1,1){#2}}\put(0,#2){\line(1,-1){#2}}
\put(#1,0){\line(-1,1){#2}}\put(#1,#2){\line(-1,-1){#2}}
\put(0,0.3){\line(1,1){0.3}}\put(1.5,0.3){\line(-1,1){0.3}}
\put(0,0){\circl}\put(0,#2){\circl}
\put(#1,0){\circl}\put(#1,#2){\circl}}

\newcommand{\SEboxxP}
{\put(0,-0.3){\LineH{2}} \put(0,0.3){\LineH{2}}
\put(0,-0.3){\LineV{0.6}} \put(2,-0.3){\LineV{0.6}}
\put(0,-0.0){\makebox(0,0){\multiput(0,0)(0,0.2){2}{\line(1,0){2}}}}
\put(0.25,0){\makebox(0,0){\multiput(0,0)(0.3,0){6}{\line(0,1){0.6}}}}
\put(0,-0.3){\circl}\put(0,0.3){\circl}
\put(2,-0.3){\circl}\put(2,0.3){\circl}}

\newcommand{\VQED}{\Photon}

\newcommand{\VQEDT}{\PhotonT}

\newcommand{\SEboxx}
{\put(0,-0.3){\LineH{2}}\put(0,0.3){\LineH{2}}
\put(0,-0.4){\multiput(0,0)(0.10,0){20}{$\cdot$}}
\put(0,-0.1){\multiput(0,0)(0.10,0){20}{$\cdot$}}
\put(0,-0.25){\multiput(0,0)(0.10,0){20}{$\cdot$}}
\put(0,0.05){\multiput(0,0)(0.10,0){20}{$\cdot$}}
\put(0,-0.3){\circl}\put(0,0.3){\circl}
\put(2,-0.3){\circl}\put(2,0.3){\circl}}

\newcommand{\SEboxxx}
{\put(0,-0.3){\LineH{3}} \put(0,0.3){\LineH{3}}
\put(0,-0.3){\LineV{0.6}} \put(3,-0.3){\LineV{0.6}}
\put(0,-0.4){\multiput(0,0)(0.10,0){30}{$\cdot$}}
\put(0,-0.1){\multiput(0,0)(0.10,0){30}{$\cdot$}}
\put(0,-0.25){\multiput(0,0)(0.10,0){30}{$\cdot$}}
\put(0,0.05){\multiput(0,0)(0.10,0){30}{$\cdot$}}
\put(0,-0.3){\circl}\put(0,0.3){\circl}
\put(3,-0.3){\circl}\put(3,0.3){\circl}}

\newcommand{\SEbox}
{\put(0,0){\makebox(0,0.35){\EEbox{1}{0.7}}} }

\newcommand{\EboxG}
{\put(0,0){\makebox(0,0.7){\EEbox{2}{0.7}}}
\put(0,0){\circl}\put(2,0){\circl}
\put(0,0.7){\circl}\put(2,0.7){\circl}}

\newcommand{\SEboxP}
{\put(0,0){\makebox(0,0){\multiput(0,0)(0,0.2){3}{\line(1,0){1}}}}
\put(0,0){\makebox(0,0){\multiput(0,0)(0.2,0){5}{\line(0,1){0.7}}}}
  \put(0,0){\makebox(0,0){\Ebox{1}{0.7}}}}

\newcommand{\photonG}
{\qbezier(0,0)(0.08333,0.125)(0.1666667,0)
\qbezier(0.1666667,0)(0.25,-0.125)(0.3333333,0)
\qbezier(0.3333333,0)(0.416667,0.125)(0.5,0)

\qbezier(0.5,0)(0.583333,-0.125)(0.666667,0)
\qbezier(0.666667,0)(0.75,0.125)(0.833333,0)
\qbezier(0.833333,0)(0.916667,-0.125)(1,0)

\qbezier(1,0)(1.083333,0.125)(1.166667,0)
\qbezier(1.166667,0)(1.25,-0.125)(1.333333,0)
\qbezier(1.333333,0)(1.416667,0.125)(1.5,0)

\qbezier(1.5,0)(1.583333,-0.125)(1.666667,0)
\qbezier(1.666667,0)(1.75,0.125)(1.833333,0)
\qbezier(1.833333,0)(1.916667,-0.125)(2,0)
\put(0,0){\circle*{0.2}}\put(2,0){\circle*{0.2}} }

\newcommand{\photonTG}[3]
{\put(0,0){\photonn{}{#2}{}}\put(1,0){\photonn{}{#2}{}}}

\newcommand{\photong}
{\qbezier(0,0)(0.08333,0.125)(0.1666667,0)
\qbezier(0.1666667,0)(0.25,-0.125)(0.3333333,0)
\qbezier(0.3333333,0)(0.416667,0.125)(0.5,0)

\qbezier(0.5,0)(0.583333,-0.125)(0.666667,0)
\qbezier(0.666667,0)(0.75,0.125)(0.833333,0)
\qbezier(0.833333,0)(0.916667,-0.125)(1,0)

\qbezier(1,0)(1.083333,0.125)(1.166667,0)
\qbezier(1.166667,0)(1.25,-0.125)(1.333333,0)
\qbezier(1.333333,0)(1.416667,0.125)(1.5,0)
\put(0,0){\circle*{0.20}}}

\newcommand{\photonnn}
{\qbezier(0,0)(0.08333,0.125)(0.1666667,0)
\qbezier(0.1666667,0)(0.25,-0.125)(0.3333333,0)
\qbezier(0.3333333,0)(0.416667,0.125)(0.5,0)

\qbezier(0.5,0)(0.583333,-0.125)(0.666667,0)
\qbezier(0.666667,0)(0.75,0.125)(0.833333,0)
\qbezier(0.833333,0)(0.916667,-0.125)(1,0)

\qbezier(1,0)(1.083333,0.125)(1.166667,0)
\qbezier(1.166667,0)(1.25,-0.125)(1.333333,0)
\qbezier(1.333333,0)(1.416667,0.125)(1.5,0) }

\newcommand{\photonGENE}[3]
{\qbezier(0,0)(0.17,-0.04)(0.2,0.1)
\qbezier(0.2,0.1)(0.23,0.32)(0.4,0.2)
\qbezier(0.4,0.2)(0.57,0.16)(0.6,0.3)
\qbezier(0.6,0.3)(0.63,0.52)(0.8,0.4)
\qbezier(0.8,0.4)(0.97,0.36)(1,0.5)
\qbezier(1,0.5)(1.03,0.72)(1.2,0.6)
\qbezier(1.2,0.6)(1.37,0.56)(1.4,0.7)
\qbezier(1.4,0.7)(1.43,0.92)(1.6,0.8)
\qbezier(1.6,0.8)(1.77,0.76)(1.8,0.9)
\qbezier(1.8,0.9)(1.83,1.12)(2,1)
\put(0,0){\circle*{0.20}} \put(2,1){\circle*{0.20}}
\put(1,0.85){\makebox(0,-0){$#1$}}
\put(-0.35,-1){\makebox(0,2){$#2$}}
\put(2.35,0){\makebox(0,2){$#3$}}}

\newcommand{\photonGWNW}[3]
{\qbezier(0,0)(-0.17,-0.04)(-0.2,0.1)
\qbezier(-0.2,0.1)(-0.23,0.32)(-0.4,0.2)
\qbezier(-0.4,0.2)(-0.57,0.16)(-0.6,0.3)
\qbezier(-0.6,0.3)(-0.63,0.52)(-0.8,0.4)
\qbezier(-0.8,0.4)(-0.97,0.36)(-1,0.5)
\qbezier(-1,0.5)(-1.03,0.72)(-1.2,0.6)
\qbezier(-1.2,0.6)(-1.37,0.56)(-1.4,0.7)
\qbezier(-1.4,0.7)(-1.43,0.92)(-1.6,0.8)
\qbezier(-1.6,0.8)(-1.77,0.76)(-1.8,0.9)
\qbezier(-1.8,0.9)(-1.83,1.12)(-2,1)
\put(0,0){\circle*{0.2}} \put(-2,1){\circle*{0.2}}
\put(-1,0.5){\makebox(0.6,0.4){$#1$}}
\put(0.35,-1){\makebox(0,2){$#3$}}
\put(-2.35,0){\makebox(0,2){$#2$}}}

\newcommand{\photonHG}
{\qbezier(0,0)(0.08333,0.125)(0.1666667,0)
\qbezier(0.1666667,0)(0.25,-0.125)(0.3333333,0)
\qbezier(0.3333333,0)(0.416667,0.125)(0.5,0)

\qbezier(0.5,0)(0.583333,-0.125)(0.666667,0)
\qbezier(0.666667,0)(0.75,0.125)(0.833333,0)
\qbezier(0.833333,0)(0.916667,-0.125)(1,0)

\qbezier(1,0)(1.083333,0.125)(1.166667,0)
\qbezier(1.166667,0)(1.25,-0.125)(1.333333,0)
\qbezier(1.333333,0)(1.416667,0.125)(1.5,0)
\put(0,0){\circle*{0.20}}\put(1.5,0){\circle*{0.20}} }

\renewcommand{\photonH}
{\photonHG\put(0.75,-0.1){\VectorR}}

\newcommand{\photonhp}
{\qbezier(0,0)(0.08333,0.125)(0.1666667,0)
\qbezier(0.1666667,0)(0.25,-0.125)(0.3333333,0)
\qbezier(0.3333333,0)(0.416667,0.125)(0.5,0)

\qbezier(0.5,0)(0.583333,-0.125)(0.666667,0)
\qbezier(0.666667,0)(0.75,0.125)(0.833333,0)
\qbezier(0.833333,0)(0.916667,-0.125)(1,0)
\put(0,0){\circle*{0.20}} \put(0.5,0){\VectorR}}

\newcommand{\photonhpL}
{\qbezier(0,0)(0.08333,0.125)(0.1666667,0)
\qbezier(0.1666667,0)(0.25,-0.125)(0.3333333,0)
\qbezier(0.3333333,0)(0.416667,0.125)(0.5,0)

\qbezier(0.5,0)(0.583333,-0.125)(0.666667,0)
\qbezier(0.666667,0)(0.75,0.125)(0.833333,0)
\qbezier(0.833333,0)(0.916667,-0.125)(1,0)
\put(1,0){\circle*{0.20}} \put(0.5,0){\Vector}}

\newcommand{\photonhpG}
{\qbezier(0,0)(0.08333,0.125)(0.1666667,0)
\qbezier(0.1666667,0)(0.25,-0.125)(0.3333333,0)
\qbezier(0.3333333,0)(0.416667,0.125)(0.5,0)

\qbezier(0.5,0)(0.583333,-0.125)(0.666667,0)
\qbezier(0.666667,0)(0.75,0.125)(0.833333,0)
\qbezier(0.833333,0)(0.916667,-0.125)(1,0)
\put(0,0){\circle*{0.20}} }

\newcommand{\photonh}
{\put(0,0){\photonhp} \put(1,0){\circle*{0.20}}}

\newcommand{\photonhG}
{\put(0,0){\photonhpG} \put(1,0){\circle*{0.20}}}

\newcommand{\photonV}[0]
{\qbezier(0,0)(-0.1666667,0.111111)(0,0.222222)
\qbezier(0,0.222222)(0.1666667,0.333333)(0,0.444444)
\qbezier(0,0.444444)(-0.1666667,0.55555)(0,0.666667)
\qbezier(0,0.666667)(0.1666667,0.777777)(0,0.888889)
\qbezier(0,0.888889)(-0.1666667,1)(0,1.111111)
\put(0,0){\circle*{0.20}} \put(0,1.111){\circle*{0.20}} }

\newcommand{\photonNEG}[3]
{\qbezier(0,0)(0.22,-0.02)(0.2,0.2)
\qbezier(0.2,0.2)(0.18,0.42)(0.4,0.4)
\qbezier(0.4,0.4)(0.62,0.38)(0.6,0.6)
\qbezier(0.6,0.6)(0.58,0.82)(0.8,0.8)
\qbezier(0.8,0.8)(1.02,0.78)(1,1)
\qbezier(1,1)(0.98,1.22)(1.2,1.2)
\qbezier(1.2,1.2)(1.42,1.18)(1.4,1.4)
\qbezier(1.4,1.4)(1.38,1.62)(1.6,1.6)
\qbezier(1.6,1.6)(1.82,1.58)(1.8,1.8)
\qbezier(1.8,1.8)(1.78,2.02)(2,2)\put(1,1){\makebox(-0.6,0.4){$#1$}}
\put(-0.35,-1){\makebox(0,2){$#2$}}
\put(2.35,1){\makebox(0,2){$#3$}}
\put(0,0){\circle*{0.20}}\put(2,2){\circle*{0.20}}
 }

\newcommand{\photonNWG}[3]
{\qbezier(0,0)(-0.22,-0.02)(-0.2,0.2)
\qbezier(-0.2,0.2)(-0.18,0.42)(-0.4,0.4)
\qbezier(-0.4,0.4)(-0.62,0.38)(-0.6,0.6)
\qbezier(-0.6,0.6)(-0.58,0.82)(-0.8,0.8)
\qbezier(-0.8,0.8)(-1.02,0.78)(-1,1) \qbezier(-1,1)
(-0.98,1.22)(-1.2,1.2) \qbezier(-1.2,1.2)(-1.42,1.18)(-1.4,1.4)
\qbezier(-1.4,1.4)(-1.38,1.62)(-1.6,1.6)
\qbezier(-1.6,1.6)(-1.82,1.58)(-1.8,1.8)
\qbezier(-1.8,1.8)(-1.78,2.02)(-2,2)
\put(-1,1){\makebox(0.4,0.7){$#1$}}
\put(-2.35,2){\makebox(0,0){$#2$}}
\put(0.35,0){\makebox(0,0){$#3$}}
}

\newcommand{\photonENEG}[3]
{\qbezier(0,0)(0.17,-0.04)(0.2,0.1)
\qbezier(0.2,0.1)(0.23,0.32)(0.4,0.2)
\qbezier(0.4,0.2)(0.57,0.16)(0.6,0.3)
\qbezier(0.6,0.3)(0.63,0.52)(0.8,0.4)
\qbezier(0.8,0.4)(0.97,0.36)(1,0.5)
\qbezier(1,0.5)(1.03,0.72)(1.2,0.6)
\qbezier(1.2,0.6)(1.37,0.56)(1.4,0.7)
\qbezier(1.4,0.7)(1.43,0.92)(1.6,0.8)
\qbezier(1.6,0.8)(1.77,0.76)(1.8,0.9)
\qbezier(1.8,0.9)(1.83,1.12)(2,1)
\put(1,0.85){\makebox(0,-0){$#1$}}
\put(-0.35,-1){\makebox(0,2){$#2$}}
\put(2.35,0){\makebox(0,2){$#3$}} \put(0,0){\circle*{0.20}}
\put(2,1){\circle*{0.20}}}

\newcommand{\photonWNWG}[3]
{\qbezier(0,0)(-0.17,-0.04)(-0.2,0.1)
\qbezier(-0.2,0.1)(-0.23,0.32)(-0.4,0.2)
\qbezier(-0.4,0.2)(-0.57,0.16)(-0.6,0.3)
\qbezier(-0.6,0.3)(-0.63,0.52)(-0.8,0.4)
\qbezier(-0.8,0.4)(-0.97,0.36)(-1,0.5)
\qbezier(-1,0.5)(-1.03,0.72)(-1.2,0.6)
\qbezier(-1.2,0.6)(-1.37,0.56)(-1.4,0.7)
\qbezier(-1.4,0.7)(-1.43,0.92)(-1.6,0.8)
\qbezier(-1.6,0.8)(-1.77,0.76)(-1.8,0.9)
\qbezier(-1.8,0.9)(-1.83,1.12)(-2,1) \put(0,0){\circle*{0.15}}
\put(-2,1){\circle*{0.15}} \put(-1,0.5){\makebox(0.6,0.4){$#1$}}
\put(0.35,-1){\makebox(0,2){$#3$}}
\put(-2.35,0){\makebox(0,2){$#2$}}}

\newcommand{\CrossphotonsG}[6]
{\put(0,0){\photonNEG{#5}{}{}} \put(2,0){\photonNWG{#6}{}{}}
\put(-0.35,0){\makebox(0,0){$#1$}}
\put(2.35,2){\makebox(0,0){$#2$}}
\put(2.35,0){\makebox(0,0){$#3$}}
\put(-0.35,2){\makebox(0,0){$#4$}}
 \put(0,0){\circle*{0.20}} \put(2,0){\circle*{0.20}}
 \put(0,2){\circle*{0.20}} \put(2,2 ){\circle*{0.20}}}

\newcommand{\photonHSG}[4]
{\qbezier(0,0)(0.08333,0.125)(0.1666667,0)
\qbezier(0.1666667,0)(0.25,-0.125)(0.3333333,0)
\qbezier(0.3333333,0)(0.416667,0.125)(0.5,0)
\qbezier(0.5,0)(0.583333,-0.125)(0.666667,0)
\qbezier(0.666667,0)(0.75,0.125)(0.833333,0)
\qbezier(0.833333,0)(0.916667,-0.125)(1,0)
\put(0.5,0.35){\makebox(0,0){$#1$}} \put(0,0){\circle*{0.20}}
\put(1,0){\circle*{0.20}}
\put(0,-0.5){\makebox(0,0){#2}}\put(1,-0.5){\makebox(0,0){#3}}}

\newcommand{\PotintG}
{\linethickness{0.3mm}\put(0,0)\dashH
\put(1.35,0){\makebox(0,0){\large$\times$}}
\put(0,0){\circle*{0.20}} }

\newcommand{\EffpotG}
{\linethickness{0.3mm}\put(0,0){\multiput(0.05,0)(0.25,0){5}{\line(1,0){0.15}}}
\put(1.55,0){\makebox(0,0){\large$\times$}}
\put(1.55,0){\circle{0.5}} \put(0,0){\circle*{0.20}} }

\newcommand{\EffpotGG}
{\linethickness{0.3mm}\put(0,0){\multiput(0.05,0)(0.25,0){5}{\line(1,0){0.15}}}
\put(1.6,0){\makebox(0,0){\large$\times$}}
\put(1.6,0){\circle{0.5}} \put(1.6,0){\circle{0.7}}
\put(0,0){\circle*{0.20}} }

\newcommand{\elstatG}[3]
{\multiput(0.06,0)(0.25,0){8}{\line(1,0){0.15}}
\put(1,0.35){\makebox(0,0){$#1$}} \put(0,0){\circle*{0.2}}
\put(2,0){\circle*{0.2}} \put(-0.35,0){\makebox(0,0){#2}}
\put(2.35,0){\makebox(0,0){#3}}}

\newcommand{\elstatHG}[3]
{\linethickness{0.3mm}\multiput(0.06,0)(0.25,0){6}{\line(1,0){0.15}}
\put(0.75,0.25){\makebox(0,0){$#1$}} \put(0,0){\circle*{0.2}}
\put(1.5,0){\circle*{0.2}} \put(-0.35,0){\makebox(0,0){#2}}
\put(2.35,0){\makebox(0,0){#3}}}

\newcommand{\VPloopG}[1]
{\put(0,0){\circle{1}}\put(0.025,-0.025){\circle{1}}\put(0.025,-0.025){\circle{1}}
\put(0.025,-0){\circle{1}}\put(0,-0.025){\circle{1}}
\put(0.85,0){\makebox(0,0){$#1$}}}

\newcommand{\LoopG}[2]
{\put(0,0){\Oval{0.6}{1.25}} \put(-0.3,0){\VectorDn}
\put(-0.65,0){\makebox(0,0){$#1$}}
\put(0.65,0){\makebox(0,0){$#2$}}}

\newcommand{\photonSEG}[5]
{\put(0,0){\photonHSG{#3}{#4}{}{}} \put(1.5,0){\VPloopD{#1}{#2}}
\put(2,0){\photonHSG{#5}{}{}{}} }

\newcommand{\GG}[2]
{\put(0,0){\LineV{#1}} \put(0,0){\circl}
\put(0,#1){\circl}\put(0,#2){\VectorUp} }

\newcommand{\Gn}[1]
{\put(0,0){\LineV{#1}} \put(0,0){\circl} \put(0,#1){\circl} }

\newcommand{\Gnn}[1]
{\put(0,0){\LineV{#1}} \put(0,#1){\circl} }

\newcommand{\GGD}[3]
{\put(0,0){\LineV{#1}} \put(0,0){\circl}
\put(0,#1){\circl}\put(0,#2){\VectorUp} \put(0,#3){\VectorUp}}

\newcommand{\GGT}[4]
{\put(0,0){\LineV{#1}} \put(0,0){\circl}
\put(0,#1){\circl}\put(0,#2){\VectorUp}
\put(0,#3){\VectorUp}\put(0,#4){\VectorUp}}

\newcommand{\GGF}[5]
{\put(0,0){\LineV{#1}} \put(0,0){\circl}
\put(0,#1){\circl}\put(0,#2){\VectorUp}
\put(0,#3){\VectorUp}\put(0,#4){\VectorUp}\put(0,#5){\VectorUp}}

\newcommand{\GGFm}[6]
{\put(0,0){\LineV{#1}} \put(0,0){\circl}
\put(0,#1){\circl}\put(0,#2){\VectorUp}
\put(0,#3){\VectorUp}\put(0,#4){\VectorUp}\put(0,#5){\VectorUp}
\put(0,#6){\VectorUp}}

\newcommand{\Gg}[1]
{\put(0,0){\LineV{#1}} \put(0,0){\circle} \put(0,#1){\circl}}

\newcommand{\GGC}[5]
{\put(0,0){\LineV{#1}} \put(0,#3){\circl}\put(0,#4){\circl}
}

\newcommand{\GGc}[3]
{\put(0,0){\LineV{#1}} \put(0,#3){\circl}
}

\newcommand{\GGUp}[0]
  {\put(0,0){\LineV{1.5}}
 }

\newcommand{\GGUpR}[0]
  {\put(0,0){\line(1,3){0.5}}
  \put(0,0){\vector(1,3){0.30}}
  }

\newcommand{\GGDnL}[0]
  {\put(0,0){\line(-1,3){0.5}}
  \put(0,0){\vector(-1,3){0.30}}
   }

\newcommand{\GGDnR}[0]
  {\put(0,0){\line(1,-3){0.5}}
  \put(0,0){\vector(1,-3){0.30}}
   }

\newcommand{\GGG}[2]
{\linethickness{0.8mm}\put(0,0){\line(0,1){#1}} \put(0,0){\circl}
\put(0,#1){\circl}\put(0.035,#2)
{\VectorUp}\put(-0.035,#2){\VectorUp} }

\newcommand{\Pot}[0]
{ \put(0,0){\photonnH}  \put(1.2,-0.15){\Large$\times$}
\put(0,0){\circl}}

\newcommand{\PotL}[0]
{ \put(0.5,0){\photonH}  \put(1.7,-0.2){\Large$\times$}
\put(-0.475,0){\VectorUp} \put(0,0){\circle{1}}}

\newcommand{\VP}[0]
{\put(0,0){\Circle{1}} 
 \put(0.1,0.5){\vector(1,0){0}}
 }

\newcommand{\VPG}[0]
{\put(2,0){\LoopT{1}} \put(0,0){\photonH}
\put(2.45,0.05){\VectorUp} }

\newcommand{\VPh}[0]
{\put(1.5,0){\LoopT{1}} \put(0,0){\photonh}
  \put(1.95,0.05){\VectorUp}}

\newcommand{\VPhh}[0]
{\put(1.5,0){\LoopT{1}} \put(0,0){\photonh}
 \put(1.5,0.475){\Vector}\put(1.5,-0.475){\VectorR}
  }

\newcommand{\VPLG}[0]
{\put(-2,0){\LoopT{1}} \put(-1.5,0){\photonH{}{}{}}
\put(-2.5,-0.05){\VectorUp}
 }

\newcommand{\VPLh}[0]
{\put(-1.5,0){\Circle{1}} \put(-1,0){\photonh}
\put(-2,-0.05){\VectorUp}
 }

\newcommand{\VPPG}[0]
{\put(0,0){\LoopTh{1.5}} \put(-0.75,0){\photonH}
\put(0,0.70){\Vector}\put(0,-0.70){\VectorR} }

\newcommand{\VPPGLab}[2]
{\put(0,0){\LoopTh{1.5}} \put(-0.75,0){\photonH}
\put(0,1.1){\makebox(0,0){$#1$}}\put(0,-1,1){\makebox(0,0){$#2$}}
\put(0,0.70){\Vector}\put(0,-0.70){\VectorR }}

\newcommand{\IrrPot}[0]
{\put(-0.1,0){\makebox(0,0){\multiput(0,0)(0.3,0){7}{$\times$}}}
 \put(0,0){\circle*{0.25}} \put(2,0){\circle*{0.25}}}

\newcommand{\IrrPots}[0]
{\put(-0.1,0){\makebox(0,0){\multiput(0,0)(0.3,0){5}{$\times$}}}
 \put(0,0){\circle*{0.15}} \put(1.5,0){\circle*{0.15}}}

\newcommand{\IrrPotS}[0]
{\put(-0.1,0){\makebox(0,0){\multiput(0,0)(0.3,0){4}{$\times$}}}
 \put(0,0){\circle*{0.25}}}

\newcommand{\LineSG}[1]
{\linethickness{0.75mm} \put(0,0){\line(1,0){#1}}
\put(0,0){\circle*{0.25}} \put(#1,0){\circle*{0.25}}}

\newcommand{\LineSS}[1]
{\linethickness{1mm} \put(0,0){\line(1,0){#1}}
\put(0,0){\circle*{0.25}} }

\newcommand{\Handle}[0]
{\put(0,0){\LoopT{1}} \put(0.5,0){\photonH}\put(2.5,0){\LoopT{1}}
\put(-0.47,-0.025){\VectorUp}\put(2.97,-0.025){\VectorUp} }

\newcommand{\Circle}[1]
{\thicklines\put(0.,0){\circle{#1}}}

\newcommand{\HandleLab}[2]
{\put(0,0){\Circle{1}}
\put(0.5,0){\photon{}{}{}}\put(3,0){\Circle{1}}
\put(-0.8,0){\makebox(0,0){$#1$}}\put(3.8,0){\makebox(0,0){$#2$}}
\put(-0.475,0){\VectorUp}\put(3.475,0){\VectorUp}}

\newcommand{\PhInt}[0]
{\put(0,0){\photon{}{}{}}\put(0,0){\vector(-1,1){0.6}}\put(2,0){\vector(1,1){0.6}}
 \put(-0.6,-0.6){\vector(1,1){0.6}}\put(2.6,-0.6){\vector(-1,1){0.6}}}

\newcommand{\PhIntS}[0]
{\put(0,0){\photong}\put(0,0){\vector(-1,1){0.6}}\put(2,0){\vector(1,1){0.6}}
 \put(-0.6,-0.6){\vector(1,1){0.6}}\put(2.6,-0.6){\vector(-1,1){0.6}}}

\newcommand{\PhIntLab}[4]
 {\put(0,0){\photon{}{}{}}\put(0,0){\vector(-1,1){0.6}}\put(2,0){\vector(1,1){0.6}}
 \put(-0.6,-0.6){\vector(1,1){0.6}}\put(2.6,-0.6){\vector(-1,1){0.6}}
 \put(-1,0.8){\makebox(0,0){#1}}\put(3,0.8){\makebox(0,0){#2}}
 \put(-1,-0.8){\makebox(0,0){#3}}\put(3,-0.8){\makebox(0,0){#4}}
 \put(0,0.3){1}\put(1.75,0.3){2}}

\newcommand{\ElSEG}[0]
{\qbezier(0,-1)(.2025,-1.1489)(0.3420,-0.9397)
\qbezier(0.3420,-0.9397)(0.4167,-0.7217)(0.6428,-0.766)
\qbezier(0.6428,-0.766)(0.8937,-0.7499)(0.866,-0.5)
\qbezier(0.866,-0.5)(0.7831,-0.2850)(0.9848,-0.1736)
\qbezier(0.9848,-0.1736)(1.1667,0)(0.9848,0.1736)
\qbezier(0.9848,0.1736)(0.7831,0.2850)(0.866,0.5)
\qbezier(0.866,0.5)(0.8937,0.7499)(0.6428,0.766)
\qbezier(0.6428,0.766)(0.4167,0.7217)(0.3420,0.9397)
\qbezier(0.3420,0.9397)(.2025,1.1489)(0,1)
\put(0,-1){\circle*{0.20}}\put(0,1){\circle*{0.20}} }

\newcommand{\ElSEg}
{\setlength{\unitlength}{0.4cm}\qbezier(0,-1)(.2025,-1.1489)(0.3420,-0.9397)
\qbezier(0.3420,-0.9397)(0.4167,-0.7217)(0.6428,-0.766)
\qbezier(0.6428,-0.766)(0.8937,-0.7499)(0.866,-0.5)
\qbezier(0.866,-0.5)(0.7831,-0.2850)(0.9848,-0.1736)
\qbezier(0.9848,-0.1736)(1.1667,0)(0.9848,0.1736)
\qbezier(0.9848,0.1736)(0.7831,0.2850)(0.866,0.5)
\qbezier(0.866,0.5)(0.8937,0.7499)(0.6428,0.766)
\qbezier(0.6428,0.766)(0.4167,0.7217)(0.3420,0.9397)
\qbezier(0.3420,0.9397)(.2025,1.1489)(0,1)
\put(0,-1){\circle*{0.225}}\put(0,1){\circle*{0.225}}}

\newcommand{\ElSEgg}
{\setlength{\unitlength}{0.8cm}\qbezier(0,-1)(.2025,-1.1489)(0.3420,-0.9397)
\qbezier(0.3420,-0.9397)(0.4167,-0.7217)(0.6428,-0.766)
\qbezier(0.6428,-0.766)(0.8937,-0.7499)(0.866,-0.5)
\qbezier(0.866,-0.5)(0.7831,-0.2850)(0.9848,-0.1736)
\qbezier(0.9848,-0.1736)(1.1667,0)(0.9848,0.1736)
\qbezier(0.9848,0.1736)(0.7831,0.2850)(0.866,0.5)
\qbezier(0.866,0.5)(0.8937,0.7499)(0.6428,0.766)
\qbezier(0.6428,0.766)(0.4167,0.7217)(0.3420,0.9397)
\qbezier(0.3420,0.9397)(.2025,1.1489)(0,1)
\put(0,-1){\circle*{0.11}}\put(0,1){\circle*{0.11}}}

\newcommand{\ElSELG}[0]
{\qbezier(0,-1)(-.2025,-1.1489)(-0.3420,-0.9397)
\qbezier(-0.3420,-0.9397)(-0.4167,-0.7217)(-0.6428,-0.766)
\qbezier(-0.6428,-0.766)(-0.8937,-0.7499)(-0.866,-0.5)
\qbezier(-0.866,-0.5)(-0.7831,-0.2850)(-0.9848,-0.1736)
\qbezier(-0.9848,-0.1736)(-1.1667,0)(-0.9848,0.1736)
\qbezier(-0.9848,0.1736)(-0.7831,0.2850)(-0.866,0.5)
\qbezier(-0.866,0.5)(-0.8937,0.7499)(-0.6428,0.766)
\qbezier(-0.6428,0.766)(-0.4167,0.7217)(-0.3420,0.9397)
\qbezier(-0.3420,0.9397)(-.2025,1.1489)(0,1)
\put(0,-1){\circle*{0.20}}\put(0,1){\circle*{0.20}} }